\documentclass{article}%
\usepackage{amsfonts}
\usepackage{amssymb}
\usepackage{amsmath}
\usepackage{graphicx}
\usepackage{caption}
\usepackage{subcaption}
\providecommand{\U}[1]{\protect\rule{.1in}{.1in}}

\begin{document}

\title{ROC Analyses Based on Measuring Evidence}
\author{Luai Al Labadi$^{\ast}$, Michael Evans$^{\ast}$ and Qiaoyu Liang$^{\ast\ast}$\\University of Toronto$^{\ast}$ and University of Waterloo$^{\ast\ast}$}
\date{}
\maketitle

\begin{abstract}
ROC analyses are considered under a variety of assumptions concerning the
distributions of a measurement $X$ in two populations. These include the
binormal model as well as nonparametric models where little is assumed about
the form of distributions. The methodology is based on a characterization of
statistical evidence which is dependent on the specification of prior
distributions for the unknown population distributions as well as for the
relevant prevalence $w$ of the disease in a given population. In all cases,
elicitation algorithms are provided to guide the selection of the priors.
Inferences are derived for the AUC as well as the cutoff $c$ used for
classification and the associated error characteristics.

\end{abstract}


\section{Introduction}

An ROC analysis is used in medical science to determine whether or not a
real-valued diagnostic $X$ for a disease or condition is useful. If the
diagnostic indicates that an individual has the condition, then this will
typically mean that a more expensive or invasive medical procedure is
undertaken. So it is important to assess the accuracy of $X.$ These methods
have a wider class of applications but our terminology will focus on the
medical context.

An approach to such analyses is presented here that is based on a
characterization of statistical evidence and which incorporates all available
information as expressed via prior probability distributions. For example,
while p-values are often used in such analyses, there are questions concerning
the validity of these quantities as characterizations of statistical evidence.
As will be seen, there are many advantages to the framework adopted here.

A common approach to the assessment of $X$ is to estimate its AUC, namely, the
probability that an individual sampled from the diseased population will have
a higher value of $X$ than an individual independently sampled from the
nondiseased population. A good $X$ should give a value of the AUC near 1 while
a value near 1/2 indicates a poor diagnostic (if the AUC is near 0, then the
classification is reversed). It is possible, however, that a diagnostic with
AUC $\approx1$ may not be suitable (see Examples 1 and 6). In particular, a
cutoff value $c$ needs to be selected so that if $X>c,$ then an individual is
classified as requiring the more invasive procedure. Inferences about the
error characteristics for the combination $(X,c),$ such as the false positive
rate, etc., are also required.

This paper is concerned with inferences about the AUC, the cutoff $c,$ and the
error characteristics. A key aspect of the analysis is the \textit{relevant
prevalence }$w.$ The phrase \textquotedblleft relevant
prevalence\textquotedblright\ means that $X$ will be applied to a certain
population, such as those patients who exhibit certain symptoms, and $w$
represents the proportion of this subpopulation who are diseased. The value of
$w$ may vary by geography, medical unit, time, etc. To make a valid assessment
of $X$ in an application, it is necessary that the information available
concerning $w$ be incorporated. This information is expressed here via an
elicited prior probability distribution for $w,$ which may be degenerate at a
single value if $w$ is known, or be quite diffuse when little is known about
$w.$ In fact, all unknown population quantities are given elicited priors.
There are many contexts where data is available relevant to the value of $w$
and this leads to a full posterior analysis for $w$ as well as for the other
quantities of interest. Even when such data is not available, however, it is
still possible to take the prior for $w$ into account so the uncertainties
concerning $w$ always play a role in the analysis.

While there are many methods available for the choice of $c,$ see
L\'{o}pez-Rat\'{o}n et al. (2014), Unal (2017), these often do not depend on
the prevalence $w$ which is a key factor in determining the true error
characteristics of $(X,c)$ in an application, see Verbakel et al. (2020). So
it is preferable to take $w$ into account when considering\ the value of a
diagnostic in a particular context. One approach to choosing $c$ is to
minimize some error criterion that depends on $w$ to obtain $c_{opt}.$ As will
be demonstrated in the examples, however, sometimes $c_{opt}$ results in a
classification that is useless.\ In such a situation a suboptimal choice of
$c$\ is required but the error characteristics can still be based on what is
known about $w$ so that these are directly relevant to the application.

Others have pointed out deficiencies in the AUC statistic and proposed
alternatives. Hand (2009) takes into account the costs associated with various
misclassification errors and argues that using the AUC is implicitly making
unrealistic assumptions concerning these costs. While costs are relevant,
costs are not incorporated here as these are often difficult to quantify. Our
goal is to express clearly what the evidence is saying about how good $(X,c)$
is via an assessment of its error characteristics. With the error
characteristics in hand, a user can decide whether or not the costs of
misclassifications are such that the diagnostic is usable. This may be a
qualitative assessment although, if numerical costs are available, these could
be subsequently incorporated. The principle here is that economic or social
factors be considered separately from what the evidence in the data says, as
it is a goal of statistics to clearly state the latter.

The framework for the analysis is Bayesian as proper priors are placed on the
unknown distribution $F_{ND}$ (the distribution of $X$ in the nondiseased
population), on $F_{D}$ (the distribution of $X$ in the diseased population)
and the prevalence $w.$ In all the problems considered, elicitation algorithms
are presented for how to choose these priors. Also, all inferences are based
on the relative belief characterization of statistical evidence where, for a
given quantity, evidence in favor (against) is obtained when posterior beliefs
are greater (less) than prior beliefs, see Evans (2015). So evidence is
determined by how the data changes beliefs. Section 2 discusses the general
framework and defines relevant quantities. Section 3 develops inferences for
these quantities for three contexts (1) $X$ is an ordered discrete variable
with no constraints on $(F_{ND},F_{D})$ (2) $X$ is a continuous variable and
$(F_{ND},F_{D})$ are normal distributions (the \textit{binormal model}) (3)
$X$ is a continuous variable and no constraints are placed on $(F_{ND}%
,F_{D}).$

There is previous work on using Bayesian methods in ROC analyses.\ For
example, Gu et al. (2008) estimate the ROC using the Bayesian bootstrap.
Carvalho et al. (2013) consider ROC analyses when there are covariates using
priors similar to those discussed in Section 3.4. Ladouceur et al. (2011) also
use priors similar to those used here but only consider the sampling regime
where the data can be used for inference about the relevant prevalence and where a
gold standard classifier is not assumed to exist.

The contributions of this paper are as follows. An elicitation algorithm is
provided for every prior used. As described in Section 2, two different
sampling regimes are considered as sampling regime (i) seems more relevant in
many medical applications than sampling regime (ii). While a prior on the
relevant prevalence is used in both sampling regimes, the posterior
distribution of this quantity is only available in sampling regime (ii) but
the prior is still used when making inferences about relevant quantities under
sampling regime (i). Inferences about the AUC, the optimal cutoff and various
error quantities associated with the cutoff are implemented for both sampling
regimes. These inferences include estimates of the AUC and the optimal cutoff
as well as exact assessments of the error in these estimates. In addition,
estimates are provided for the error characteristics of the classification, at the cutoff used, that
determine the value of the diagnostic in an application. It is shown that
sometimes a useful optimal cutoff does not exist so some other choice is
necessary. In each case the hypothesis $H_{0}:$ AUC $>1/2$ is first assessed
and if evidence is found in favor of this, the prior is then conditioned on
this event being true for inferences about the remaining quantities. Three
contexts are considered, the diagnostic takes finitely many values, the
diagnostic is normally distributed and the diagnostic is continuous but not normal. A
thorough analysis is made of the binormal model and it is shown that, unless
certain conditions on the model parameters are satisfied, then a useful
optimal cutoff is not available. Hypothesis assessments are made to determine
if these conditions hold. Based on the binormal model, a nonparametric Bayes
model is developed that allows for deviation from normality.

\section{The Problem}

Consider the formulation of the problem as presented in Obuchowski and Bullen
(2018), Zhou et al. (2011) but with somewhat different notation. There is a
measurement $X:\Omega\rightarrow R^{1}$ defined on a population $\Omega
=\Omega_{D}\cup\Omega_{ND},$ with $\Omega_{D}\cap\Omega_{ND}=\phi,$ where
$\Omega_{D}$ is comprised of those with a particular disease, and $\Omega
_{ND}$ represents those without the disease. So $F_{ND}(c)=\#(\{\omega
\in\Omega_{ND}:X(\omega)\leq c\})/\#(\Omega_{ND})\ $is the conditional cdf of
$X$ in the nondiseased population, and $F_{D}(x)=\#(\{\omega\in\Omega
_{D}:X(\omega)\leq x\})/\#(\Omega_{D})$ is the conditional cdf of $X$ in the
diseased population. It is assumed that there is a gold standard classifier,
typically much more difficult to use than $X,$ such that for any $\omega
\in\Omega$ it can be determined definitively if $\omega\in\Omega_{D}$ or
$\omega\in\Omega_{ND}.$ There are two ways in which one can sample from
$\Omega,$ namely,

\begin{quote}
(i) take samples from each of $\Omega_{D}$ and $\Omega_{ND}$ separately or

(ii) take a sample from $\Omega.$
\end{quote}

\noindent The sampling method used affects the inferences that can be drawn
and for many studies (i) is the relevant sampling mode.

It supposed that the greater the value $X(\omega)$ is for individual $\omega,$
the more likely it is that $\omega\in\Omega_{D}.$ For the classification, a
cutoff value $c$ is required such that, if $X(\omega)>c$, then $\omega$ is
classified as being in $\Omega_{D}$ and otherwise is classified as being in
$\Omega_{ND}.$ But $X$ is an imperfect classifier for any $c$ and it is
necessary to assess the performance of $(X,c)$. It seems natural that a value
of $c$ be used that is optimal in some sense related to the error
characteristics of this classification. Table \ref{table 1} gives the relevant
probabilities for classification into $\Omega_{D}$ and $\Omega_{ND}$, together
with some common terminology, in a \textit{confusion matrix}.%

\begin{table}[tbp] \centering
\begin{tabular}
[c]{|c|c|c|}\hline
& $\Omega_{D}$ & $\Omega_{ND}$\\\hline
$X>c$ & $%
\begin{array}
[c]{c}%
\text{TPR}(c)=1-F_{D}(c)\\
\text{\textit{sensitivity} (\textit{recall}) or}\\
\text{true positive rate}%
\end{array}
$ & $%
\begin{array}
[c]{c}%
\text{FPR}(c)=1-F_{ND}(c)\\
\text{false positive rate}%
\end{array}
$\\\hline
$X\leq c$ & $%
\begin{array}
[c]{c}%
\text{FNR}(c)=F_{D}(c)\\
\text{false negative rate}%
\end{array}
$ & $%
\begin{array}
[c]{c}%
\text{TNR}(c)=F_{ND}(c)\\
\text{\textit{specificity} or}\\
\text{true negative rate}%
\end{array}
$\\\hline
\end{tabular}
\caption{Error probabilities when $X>c$  indicates a positive
.}\label{table 1}%
\end{table}%

Another key ingredient is the prevalence $w=\#(\Omega_{D})/\#(\Omega)$ of the
disease in $\Omega$. In practical situations, it is necessary to also take $w$
into account in assessing the error in $(X,c).$ The following error
characteristics depend on $w,$%
\begin{align*}
\text{Error}(c)  &  =\text{misclassification rate }=w\text{FNR}%
(c)+(1-w)\text{FPR}(c),\\
\text{FDR}(c)  &  =\text{false discovery rate}=\frac{(1-w)\text{FPR}%
(c)}{w(1-\text{FNR}(c))+(1-w)\text{FPR}(c)},\\
\text{FNDR}(c)  &  =\text{false nondiscovery rate}=\frac{w\text{FNR}%
(c)}{w\text{FNR}(c)+(1-w)(1-\text{FPR}(c))}.
\end{align*}
Under sampling regime (ii) and cutoff $c,$ Error$(c)$ is the probability of
making an error, FDR$(c)$ is the conditional probability of misclassifying a
subject as positive given that it is classified as positive and FNDR$(c)$ is
the conditional probability of misclassifying a subject as negative given that
it is classified as negative. It is often observed that when $w$ is very small
and FNR$(c)$ and FPR$(c)$ are small, then FDR$(c)$ can be big. This is
sometimes referred to as the \textit{base rate fallacy }as, even though the
test appears to be a good one, there is a high probability that an individual
classified as having the disease will be misclassified. For example, if $w=$
FNR$(c)=$ FPR$(c)=0.05,$ then Error$(c)=.05$, FDR$(c)=0.50,$
FNDR$(c)=2.76\times10^{-3}$ and when $w=0.01,$ then Error$(c)=.05$,
FDR$(c)=0.84,$ FNDR$(c)=5.31\times10^{-4}.$ In these cases the false
nondiscovery rate is quite small while the false discovery rate is large. If
the disease is highly contagious, then these probabilities may be considered
acceptable but indeed they need to be estimated. Similarly, FNDR$(c)$ may be
small when FNR$(c)$ is large and $w$ is very small.

It is naturally desirable to make inference about an optimal cutoff $c_{opt}$
and its associated error quantities. For a given value of $w,$ the optimal
cutoff will be defined here as $c_{opt}=\arg\inf$ Error$(c)$, the value which
minimizes the probability of making an error. Other choices for determining a
$c_{opt}$ can be made, and the analysis and computations will be quite
similar, but our thesis is that, when possible, any such criterion should
involve the prior distribution of the relevant prevalence $w.$ As demonstrated
in Example 6 this can sometimes lead to useless values of $c_{opt}$ even when
the AUC is large. While this situation calls into question the value of the
diagnostic, a suboptimal choice of $c$ can still be made according to some
alternative methodology like the use of Youden's index (maximizing
$1-2$Error$(c)$ over $c$ with $w=1/2$). The methodology developed here provides an estimate
of the $c$ to be used$,$ together with an exact assessment of the error in this
estimate, as well as providing estimates of the associated error characteristics of the classification.

\subsection{The AUC and ROC}

Consider two situations where $F_{ND},F_{D}$ are either both absolutely
continuous or both discrete. In the discrete case, suppose that these
distributions are concentrated on a set of points $c_{1}<c_{2}<\cdots<c_{m}.$
When $\omega_{D},\omega_{ND}$ are selected using sampling scheme (i), then the
probability that a higher score is received on diagnostic $X$ by a diseased
individual than a nondiseased individual is%
\begin{equation}
\text{AUC}=\left\{
\begin{tabular}
[c]{ll}%
$\int_{-\infty}^{\infty}(1-F_{D}(c))\,f_{ND}(c)\,dc\smallskip$ & abs. cont.\\
$\sum_{i=1}^{m}(1-F_{D}(c_{i}))(F_{ND}(c_{i})-F_{ND}(c_{i-1}))$ & discrete.
\end{tabular}
\ \ \ \ \ \ \ \ \ \ \ \ \ \ \ \ \ \ \right.  \label{auc}%
\end{equation}
Under the assumption that $F_{D}(c)$ is constant on $\{c:F_{ND}(c)=p\}$ for
every $p\in\lbrack0,1],$ there is a function ROC (\textit{receiver operator
curve}) such that $1-F_{D}(c)=$ ROC$(1-F_{ND}(c))$ so AUC$=\int_{-\infty
}^{\infty}$ROC$(1-F_{ND}(c))\,F_{ND}(dx).$ Putting $p=1-F_{ND}(c),$ then
ROC$(p)=1-F_{D}(F_{ND}^{-1}(1-p)).\ $In the absolutely continuous case,
AUC$=\int_{0}^{1}$ROC$(p)\,dp$ which is the \textit{area under the curve}
given by the ROC function. The area under the curve interpretation is
geometrically evocative but is not necessary for (\ref{auc}) to be meaningful.

It is commonly suggested that a good diagnostic $X$ will have AUC close to 1
while a value close to 1/2 suggests a poor diagnostic. It is surely the case,
however, that the utility of $X$ in practice will depend on the cutoff $c$
chosen and the various error characteristics associated with this choice. So
while the AUC can be used to screen diagnostics, it is only part of the
analysis and inferences about the error characteristics are required to truly
assess the performance of a diagnostic. Consider an example.\smallskip

\noindent\textbf{Example 1.} Suppose that $F_{D}=F_{ND}^{q}$ for some $q>1,$
where $F_{ND}$ is continuous, strictly increasing with associated density
$f_{ND}.$ Then using (\ref{auc}), AUC $=1-1/(q+1)$ which is approximately 1
when $q$ is large. The optimal $c$ minimizes Error$(c)=wF_{ND}^{q}%
(c)+(1-w)(1-F_{ND}(c))$ which implies $c$ satisfies $F_{ND}%
(c)=\{(1-w)/qw\}^{1/(q-1)}$ when $q>(1-w)/w$ and the optimal $c$ is otherwise
$c=\infty$. If $q=99,$ then AUC $=0.99$ and with $w=0.025,(1-w)/w=39<q$ so
FNR$(c_{opt})=0.390,$ FPR$(c_{opt})=0.009,$ Error$(c_{opt})=0.019,$
FDR$(c_{opt})=0.009$ and FNDR$(c_{opt})=0.010$. So $X$ seems like a good
diagnostic via the AUC and the error characteristics that depend on the
prevalence although within the diseased population the probability is $0.39$
of not detecting the disease. If instead $w=0.01,$ then the AUC is the same
but $q=99=(1-w)/w$ and the optimal classification always classifies an
individual as non-diseased which is useless. So the AUC does not indicate
enough about the characteristics of the diagnostic to determine if it is
useful or not. It is necessary to look at the error characteristics of the classification at the
cutoff value that will actually be\ used, to determine if a diagnostic is
suitable and this implies that information about $w$ is necessary in an
application. $\blacksquare$

\section{Inference}

Suppose we have a sample of $n_{D}$ from $\Omega_{D}$, namely, $x_{D}%
=(x_{D1},\ldots,x_{Dn_{D}})$ and a sample of $n_{ND}$ from $\Omega_{ND}$,
namely, $x_{ND}=(x_{ND1},\ldots,x_{NDn_{ND}})$ and the goal is to make
inference about the AUC, some cutoff $c$ and the error characteristics
FNR$(c),$ FPR$(c),$ Error$(c),$ FDR$(c)$ and FNDR$(c)$. For the AUC it makes
sense to first assess the hypothesis $H_{0}:$ AUC $>1/2$ via stating whether
there is evidence for or against $H_{0}$ together with an assessment of the
strength of this evidence. Estimates are required for all of these quantities,
together with\ an assessment of the accuracy of the estimate.

As stated in the Introduction, several different contexts are considered and
the approach here is Bayesian with a prior placed on $(F_{ND},F_{D})$ as well
as the relevant prevalence $w.$ The specific inferences are derived via the
\textit{principle of evidence}: if the posterior probability of an event is
greater (smaller) than the prior probability of the event, then there is
evidence in favor of (against) the event being true. This approach is
implemented via the relative belief ratio (see Evans (2015)) which is
effectively the ratio of the posterior probability to the prior probability of
the event in question. So if the relative belief ratio is greater than (less
than) 1 there is evidence in favor of (against) the event being true.

\subsection{The prevalence}

Consider first inferences for the relevant prevalence $w.$ If $w$ is known
then nothing further needs to be done but otherwise this quantity needs to be
taken into account when assessing the value of the diagnostic and so
uncertainty about $w$ needs to be addressed.

If the full data set is based on sampling scheme (ii), then $n_{D}\sim$
binomial$(n,w).$ A natural prior $\pi_{W}$ to place on $w$ is a beta$(\alpha
_{1w},\alpha_{2w})$ distribution$.$ The hyperparameters are chosen based on
the elicitation algorithm discussed in Evans et al. (2017) where interval
$[l,u]$ is chosen such that it is believed that $w\in\lbrack l,u]$ with prior
probability $\gamma.$ Here $[l,u]$ is chosen so that we are virtually certain
that $w\in\lbrack l,u]$ and $\gamma=0.99$ then seems like a reasonable choice.
Note that choosing $l=u$ corresponds to $w$ being known and so $\gamma=1$ in
that case. Next pick a point $\xi_{w}\in\lbrack l,u]$ for the mode of the
prior and a reasonable choice might be $\xi_{w}=(l+u)/2.$ Then putting
$\tau_{w}=\alpha_{1w}+\alpha_{2w}-2$ leads to the parameterization
beta$(\alpha_{1w},\alpha_{2w})=$ beta$(1+\tau_{w}\xi_{w},1+\tau_{w}(1-\xi
_{w}))$ where $\xi_{w}$ locates the mode and $\tau_{w}$ controls the spread of
the distribution about $\xi_{w}.$ Here $\tau_{w}=0$ gives the uniform
distribution and $\tau_{w}=\infty$ gives the distribution degenerate at
$\xi_{w}.$ With $\xi_{w}$ specified, $\tau_{w}$ is the smallest value of
$\tau_{w}$ such that the probability content of $[l,u]$ is $\gamma$ and this
is found iteratively. For example, if $[l,u]=[0.60,0.70]$ and $\gamma=0.99,$
so $w$ is known reasonably well, then $\xi_{w}=(l+u)/2=0.65$ and $\tau
_{w}=601.1,$ so the prior is beta$(391.72,211.39)$ and the posterior is
beta$(391.72+n_{D},211.39+n_{ND}).$

The estimate of $w$ is then obtained by maximizing the relative belief ratio
$RB(w\,|\,n_{D},n_{ND})=\pi_{W}(w\,|\,n_{D},n_{ND})/\pi_{W}(w),$ the ratio of
the posterior to the prior, as this value has the most evidence in its favor.
In this case the estimate is the MLE, namely, $w(n_{D},n_{ND})=n_{D}%
/(n_{D}+n_{ND}).$ The accuracy of this estimate is measured by the size of the
plausible region $Pl(n_{D},n_{ND})=\{w:RB(w\,|\,n_{D},n_{ND})>1\},$ the set of
all $w$ values for which there is evidence in favor. For example, if $n=100$
and $n_{D}=68,$ then $w(68,32)=0.68$ and $Pl(68,32)=[0.647,0.712]$ which has
posterior content $0.651.$ So the data suggest that the upper bound of
$u=0.70$ is too strong although the posterior belief in this interval is not
very high.

The prior and posterior distributions of $w$ play a role in inferences about
all the quantities that depend on the prevalence. In the case where the cutoff
is determined by minimizing the probability of a misclassification, then
$c_{opt},$ FNR$(c_{opt}),$ FPR$(c_{opt}),$ Error$(c_{opt}),$ FDR$(c_{opt})$
and FNDR$(c_{opt})$ all depend on the prevalence. Under sampling scheme (i),
however, only the prior on $w$ has any influence when considering the
effectiveness of $X.$ Inference for these quantities is now discussed in both cases.

\subsection{Ordered discrete diagnostic}

Suppose $X$ takes values on the finite ordered scale $c_{1}<c_{2}<\cdots
<c_{m}$ and let $p_{NDi}=P(X(\omega_{ND})=c_{i}),p_{Di}=P(X(\omega_{D}%
)=c_{i})$ so $F_{ND}(c_{i})=%
{\textstyle\sum_{j=1}^{i}}
p_{NDj}$ and $F_{D}(c_{i})=%
{\textstyle\sum_{j=1}^{i}}
p_{Dj}$. These imply that FPR$(c_{i})=1-%
{\textstyle\sum_{j=1}^{i}}
p_{NDi},$ FNR$(c_{i})=%
{\textstyle\sum_{j=1}^{i}}
p_{Di},$ AUC$(p_{ND},p_{D})=\sum_{i=1}^{m}\left(  1-\text{FNR}(c_{i})\right)
p_{NDi}$ with the remaining quantities defined similarly. Evans et al. (2017)
can be used to obtain independent elicited Dirichlet priors%
\begin{equation}
p_{ND}\sim\text{Dirichlet}(\alpha_{ND1},\ldots,\alpha_{NDm}),\ p_{D}%
\sim\text{Dirichlet}(\alpha_{D1},\ldots,\alpha_{Dm}) \label{prior}%
\end{equation}
on these probabilities by placing either upper or lower bounds on each cell
probability that hold with virtual certainty $\gamma,$ as discussed for the
beta prior on the prevalence. If little information is available, it is
reasonable to use uniform (Dirichlet$(1,\ldots,1)$) priors on $p_{ND}$ and
$p_{D}.$ This together with the independent prior on $w$ leads to prior
distributions for the AUC, $c_{opt}$ and all the quantities associated with
error assessment such as FNR$(c_{opt}),$ etc.

The data $(x_{D},x_{ND})$ give the counts $f_{ND}=(f_{ND1},\ldots,f_{NDm})$
and $f_{D}=(f_{D1},\ldots,f_{Dm})$ which in turn lead to the independent
posteriors
\begin{equation}
p_{ND}\,|\,f_{ND}\sim\text{Dirichlet}(\alpha_{ND}+f_{ND}),\text{ }%
p_{D}\,|\,f_{D}\sim\text{Dirichlet}(\alpha_{D}+f_{D}). \label{post}%
\end{equation}
Under sampling regime (ii) this, together with the independent posterior on
$w,$ leads to posterior distributions for all the quantities of interest.
Under sampling regime (i), however, the logical thing to do, so the inferences
reflect the uncertainty about $w,$ is to only use the prior on $w$ when
deriving inferences about any quantities that depend on this such as $c_{opt}$
and the various error assessments.

Consider inferences for the AUC. The first inference should be to assess the
hypothesis $H_{0}:$ AUC $>1/2$ for, if $H_{0}$ is false, then $X$ would seem
to have no value as a diagnostic (the possibility that the directionality is
wrong is ignored here). The relative belief ratio $RB(H_{0}\,|\,f_{ND}%
,f_{ND})=\Pi(H_{0}\,|\,f_{ND},f_{ND})/\Pi(H_{0})$ is computed and compared to
1. If it is concluded that $H_{0}$ is true, then perhaps the next inference of
interest is to estimate the AUC via the relative belief estimate. The prior
and posterior densities of the AUC are not available in closed form so
estimates are required and density histograms are employed here for this. The
set $(0,1]$ is discretized into $L$ subintervals $(0,1]=\cup_{i=1}^{L}(\left(
i-1)/L,i/L\right]  ,$ and putting $a_{i}=(i-1/2)/L,$ the value of the prior
density $p_{\text{AUC}}(a_{i})$ is estimated by $L \times ($proportion of prior
simulated values of AUC in $(i-1,i]/L)$ and similarly for the posterior
density $p_{\text{AUC}}(a_{i}\,|\,f_{ND},f_{D}).$ Then $RB_{\text{AUC}%
}(a\,|\,f_{ND},f_{ND})$ is maximized to obtain the relative belief estimate
AUC$(f_{ND},f_{D})$ together with the plausible region and its posterior
content. These quantities are obtained for $c_{opt}$ in a similar fashion,
although $c_{opt}$ has prior and posterior distribution concentrated on
$\{c_{1},c_{2},\ldots,c_{m}\}$ so there is no need to discretize$.$ For the
estimate $c_{opt}(f_{ND},f_{D})$, estimates of FNR$(c_{opt}(f_{ND},f_{D})),$
FPR$(c_{opt}(f_{ND},f_{D})),$ Error$(c_{opt}(f_{ND},f_{D})),$ FDR$(c_{opt}%
(f_{ND},f_{D}))$ and FNDR$(c_{opt}(f_{ND},f_{D}))$ are obtained as these
indicate the performance of the diagnostic in practice. The relative belief
estimates of these quantities are easily obtained in a second simulation where
$c_{opt}(f_{ND},f_{D})$ is fixed.

Consider now an example.\smallskip

\noindent\textbf{Example 2. }\textit{Simulated example.}

For $k=5,$ data was generated as
\begin{align*}
f_{ND}  &  \sim\text{multinomial}(50,0.5,0.2,0.1,0.1,0.1)\text{ obtaining
}f_{ND}=(29,7,4,5,5),\\
f_{D}  &  \sim\text{multinomial}(100,0.1,0.1,0.2,0.3,0.3)\text{ obtaining
}f_{D}=(14,7,25,33,21).
\end{align*}
With these choices for $p_{ND},p_{D}$ the true values are\ AUC$=0.65$ and with
$w=0.65,$ $c_{opt}=2,$ FNR$(c_{opt})=0.200,$ FPR$(c_{opt})=0.300,$
Error$_{w}(c_{opt})=0.235,$ FDR$(c_{opt})=0.168$ and FNDR$(c_{opt})=0.347$. So
$X$ is not an outstanding diagnostic but with these error characteristics it
may prove suitable for a given application. Uniform, namely,
Dirichlet$(1,1,1,1,1),$ priors were placed on $p_{ND}\ $and $p_{D},$
reflecting little knowledge about these quantities.

\begin{figure}%
\centering
\begin{tabular}{@{}cc@{}}
\includegraphics[width=6cm]{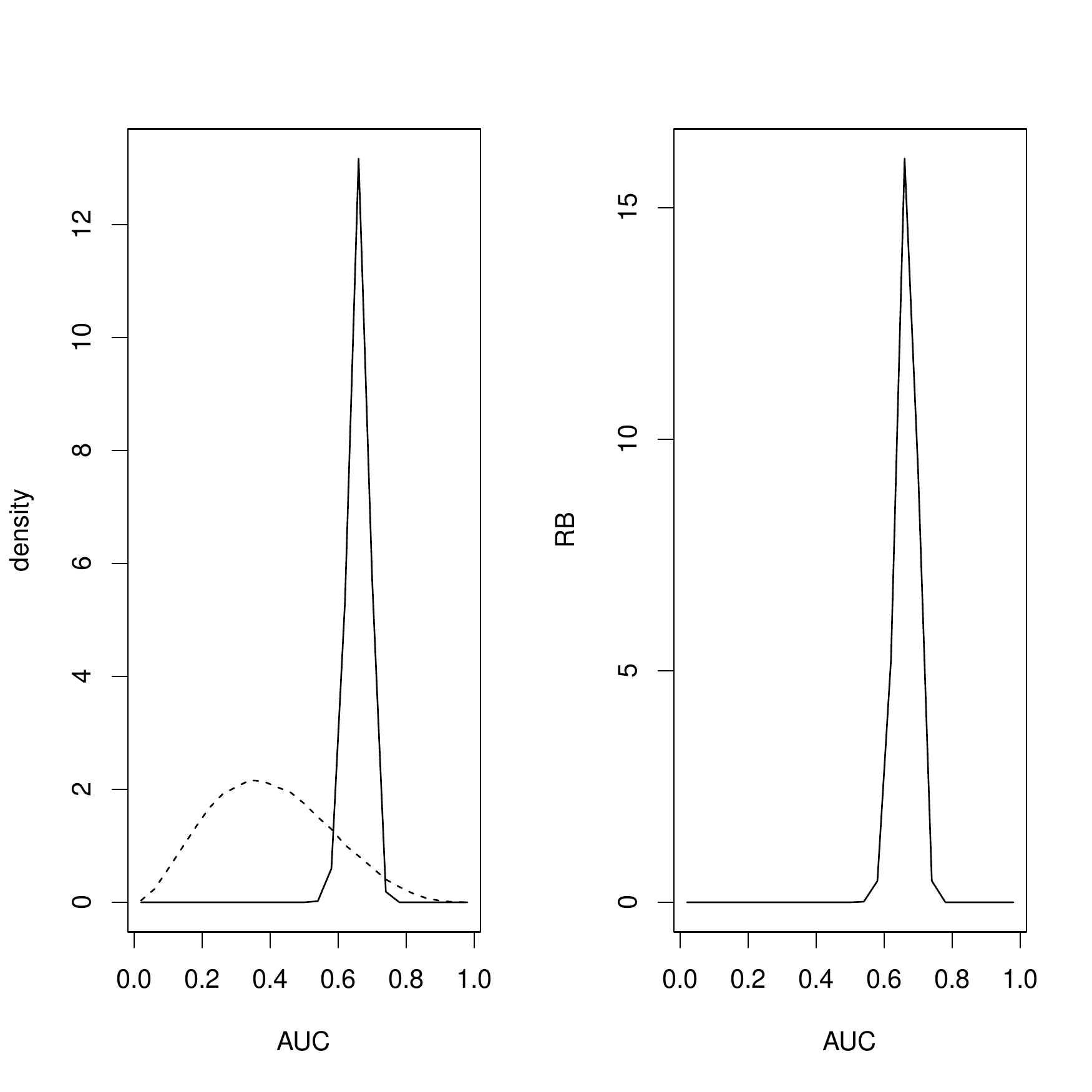} &
\includegraphics[width=6cm]{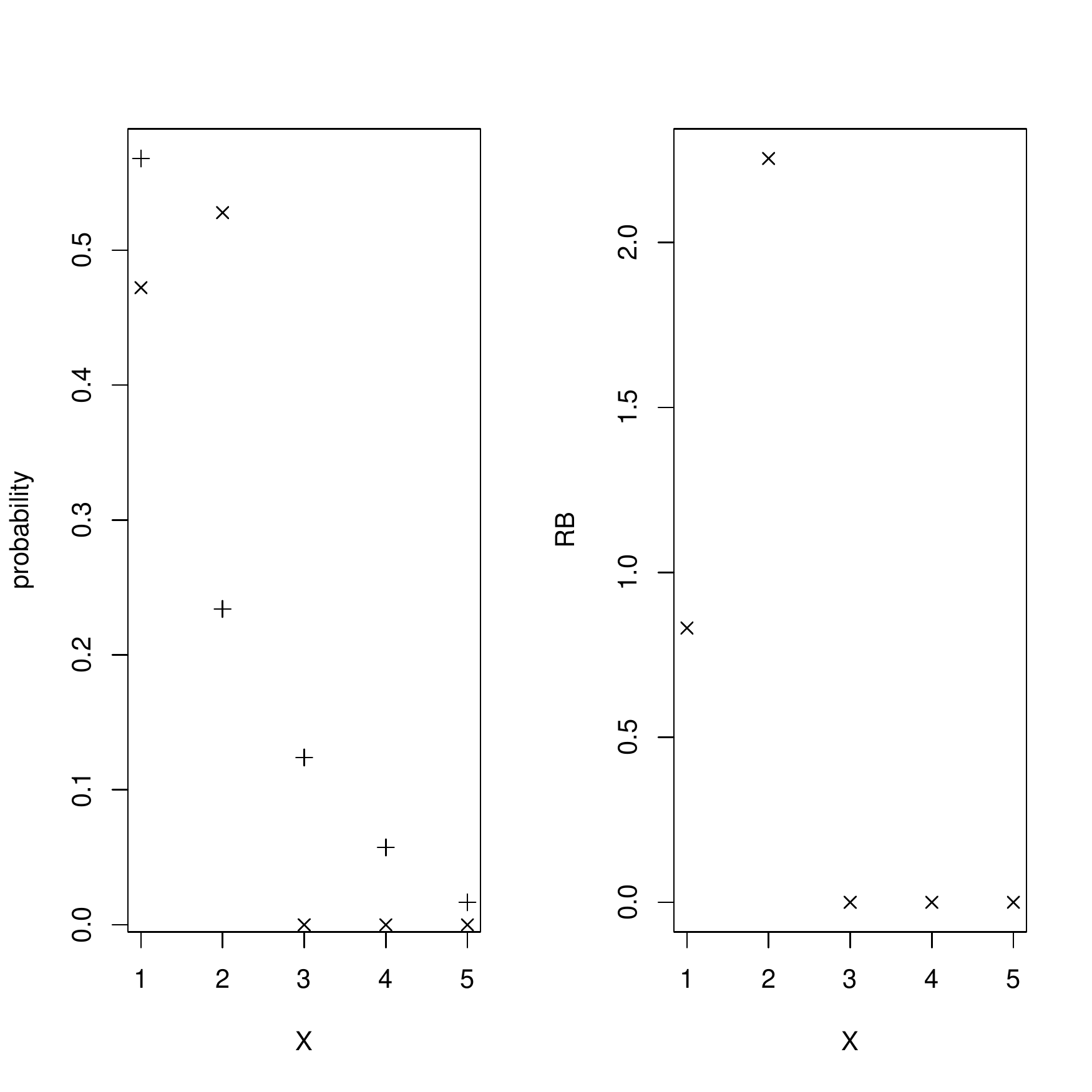} \\
\end{tabular}
\caption{In Example 2, plots of the prior (- - -), the posterior (---) and the 
RB ratio of the AUC in the left two panes and plots of the the prior (+),  the 
posterior ($\times$) and the RB ratio of $c_{opt}$ in the right two panes.}\label{twofig}%
\end{figure}

Simulations based on\ Monte Carlo sample sizes of $N=10^{5}$ from the prior
and posterior distributions of $p_{ND}\ $and $p_{D}$ were conducted and the
prior and posterior distributions of the quantities of interest obtained. The
hypothesis $H_{0}:$ AUC $>0.5$ is assessed by $RB_{\text{AUC}}%
((0.50,1.00]\,|\,f_{ND},f_{D})=3.15.$ So there is evidence in favor of $H_{0}$
and the strength of this evidence is measured by the posterior probability
content of $(0.50,1.00]$ which equals $1.0$ to machine accuracy and so this is
categorical evidence in favor of $H_{0}.$ For the continuous quantities a grid
based on $L+1=25$ equispaced points $\{0,0.04,0.08,\ldots,1.00\}$ was used and
all the mass in the interval $(i-1,i]/L$ assigned to the midpoint $(i-1/2)/L.$
Figure  \ref{twofig} contains plots of the prior and posterior densities and
relative belief ratio of the AUC. The relative belief estimate of the AUC is
AUC$(f_{ND},f_{D})=0.66$ with $Pl_{\text{AUC}}(f_{ND},f_{D})=[0.60,0.72]$
having posterior content $0.97.$ Certainly a finer partition of $[0,1]$ than
just 24 intervals is possible, but even in this relatively coarse case the
results are quite accurate.

Supposing that the relevant prevalence is known to be $w=0.65,$ Figure
\ref{twofig} contains plots of the prior and posterior densities and relative
belief ratio of $c_{opt}.$ The relative belief estimate is $c_{opt}%
(f_{ND},f_{D})=2$ with $Pl_{c_{opt}}(f_{ND},f_{D})=\{2\}$ with posterior
probability content $0.53$ so the correct optimal cut-off has been identified
but there is a degree of uncertainty concerning this. The error
characteristics that tell us about the utility of $X$ as a diagnostic are
given by the relative belief estimates (column (a)) in Table \ref{table4}. It
is interesting to note that the estimate of Error$(c_{opt})$ is determined by
the prior and posterior distributions of a convex combination of
FPR$(c_{opt})$ and FNR$(c_{opt})$ and the estimate is not the same convex
combination of the estimates of FPR$(c_{opt})$ and FNR$(c_{opt})$. So, in this
case Error$(c_{opt})$ seems like a much better assessment of the performance
of the diagnostic.%

Suppose now that the prevalence is not known but there is a beta$(1+\tau
_{w}\xi_{w},1+\tau_{w}(1-\xi_{w}))$ prior specified for $w$ and consider the
choice discussed in Section 3.1 where $\xi_{w}=0.65$ and $\tau_{w}=601.1.$
When the data is produced according to sampling regime (i), then there is no
posterior for $w$ but this prior can still be used in determining the prior
and posterior distributions of $c_{opt}$ and the associated error
characteristics. When this simulation was carried out $c_{opt}(f_{ND}%
,f_{D})=2$ with $Pl_{c_{opt}}(f_{ND},f_{D})=\{2\}$ with posterior probability
content $0.53.$ and column (b) of Table \ref{table4} gives the estimates of
the error characteristics. So other than the estimate of the FPR, the results
are similar. Finally, assuming that the data arose under sampling scheme (ii),
then $w$ has a posterior distribution and using this gives $c_{opt}%
(f_{ND},f_{D})=2$ with $Pl_{c_{opt}}(f_{ND},f_{D})=\{2\}$ with posterior
probability content $0.52$ and error characteristics as in column (c) of Table
\ref{table4}$.$ These results are the same as if the prevalence is known which
is sensible as the posterior concentrates about the true value more than the prior.%

\begin{table}[tb] \centering
\begin{tabular}
[c]{|l|c|c|c|}\hline
Quantity & Estimate (a) & Estimate (b) & Estimate (c)\\\hline
$\text{FPR}(c_{opt})$ & $0.30$ & $0.26$ & $0.30$\\
$\text{FNR}(c_{opt})$ & $0.22$ & $0.22$ & $0.22$\\
$\text{Error}(c_{opt})$ & $0.22$ & $0.22$ & $0.22$\\
$\text{FDR}(c_{opt})$ & $0.14$ & $0.14$ & $0.14$\\
$\text{FNDR}(c_{opt})$ & $0.34$ & $0.34$ & $0.34$\\\cline{1-2}\cline{2-4}%
\cline{3-4}%
\end{tabular}
\caption{The estimates of the error characteristcs of $X$ at $ c_{opt} =2 $ in Example 2 where (a) $w$ is assumed known, (b)  only the prior for $w$ is available, (c)  the posterior for $w$ is also available.}\label{table4}%
\end{table}%

\begin{figure}
[t]
\begin{center}
\includegraphics[width=6cm]%
{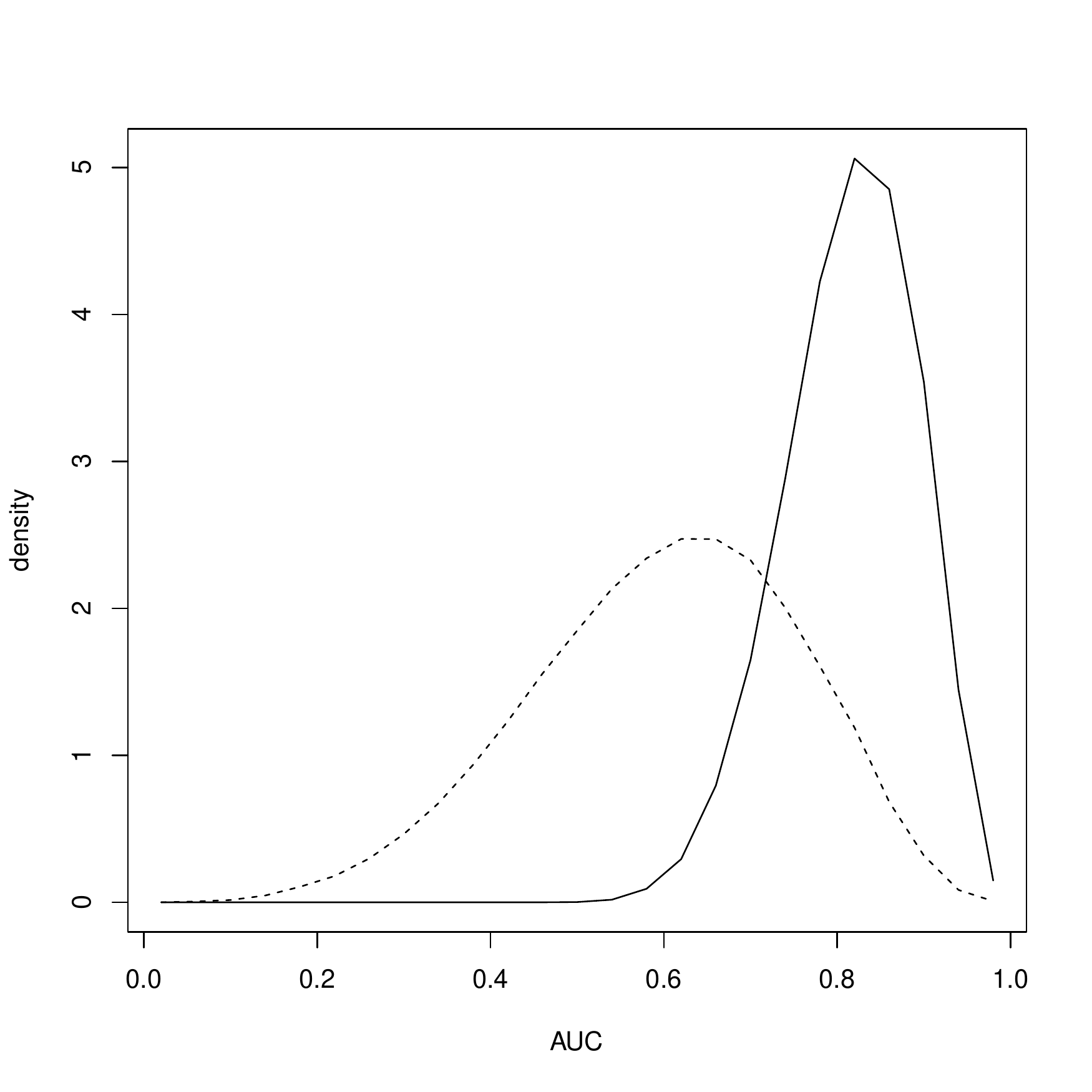}%
\caption{Prior density of the AUC when $p_{D}$ is uniform on set of
nondecreasing probabilities independent of $p_{ND}$ uniform on the set of
nonincreasing probabilities (--) as well when $p_{D}$ is uniformly distributed
on the set of nodecreasing probabilities independent of $p_{ND}$ uniform on
$S_{k}$ (- -).}%
\label{plot4}%
\end{center}
\end{figure}

Another somewhat anomalous feature of this example is the fact that uniform
priors on $p_{D}$ and $p_{ND}$ do not lead to a prior on the AUC that is even
close to uniform. In fact, these choices put more weight against
a diagnostic with AUC $>1/2$ and indeed most choices of $p_{D}$ and
$p_{ND}$ will not satisfy this. Another possibility is to require $p_{ND1}%
\geq\cdots\geq p_{ND1}$ and $p_{D1}\leq\cdots\leq p_{D1},$ namely, require
monotonicity of the probabilities. A result in\ Englert et al. (2018) implies
that $p_{ND}$ satisfies this iff $p_{ND}=A_{k}p_{*ND}$ where $p
_{*ND}\in S_{k},$ the standard $(k-1)$-dimensional simplex, and $A_{k}\in
R^{k\times k}$ with $i$-ith row equal to $(0,\ldots,0,1/i,1/(i+1),\ldots,1/k)$
and $p_{D}$ satisfies this iff $p_{D}=B_{k}p_{*D}$ where $p_{*D}\in
S_{k}$ and $B_{k}=I_{k}^{\ast}A_{k}$ where $I_{k}^{\ast}$ $\in R^{k\times k}$
contains all 0's except for 1's on the crossdiagonal. If $p_{*ND}$ and
$p_{*D}$ are independent and uniform on $S_{k},$ then $p_{D}$ and $p_{ND}$
are independent and uniform on the sets of probabilities satisfying the
corresponding monotonicities and Figure \ref{plot4} has a plot of the prior of
the AUC when this is the case. It is seen that this prior puts most of its weight in favor
of AUC $>1/2.$ Figure \ref{plot4} also has a plot the prior of the AUC when
$p_{D}$ is uniform on the set of all nondecreasing probabilities and $p_{ND}$
is uniform on $S_{k}.$ This reflects a much more modest belief that $X$ will
satisfy AUC $>1/2$ and indeed this may be a more appropriate prior than using
uniform distributions on $S_{k}.$ Englert et al. (2018) also provides
elicitation algorithms for choosing alternative Dirichlet distributions for
$p_{*ND}$ and $p_{*D}.$ $\blacksquare$%

When $H_{0}:$ AUC $>0.5$ is accepted, it makes sense to use the conditional
prior, given that this event is true, in the inferences. As such it is
necessary to condition the prior on the event $\sum_{i=1}^{m}\left(
\sum_{j=1}^{i}p_{Dj}\right)  p_{NDi}\leq1/2.$ In general, it isn't clear how
to generate from this conditional prior but depending on the size of $m$ and
the prior, a brute force approach is to simply generate from the unconditional
prior and select those samples for which the condition is satisfied and the
same approach works with the posterior.\smallskip

\noindent\textbf{Example 2. }\textit{Simulated example (continued).}

Here $m=5,$ and using uniform priors for $p_{ND}$ and $p_{D}$, the prior
probability of AUC\ $>0.5$ is $0.281\ $while the posterior probability is
$0.998$ so the posterior sampling is much more efficient. Choosing priors that
are more favorable to AUC\ $>0.5$ will improve the efficiency of the prior
sampling. Using the conditional priors led to AUC$(f_{ND},f_{D})=0.66$ with
$Pl_{\text{AUC}}(f_{ND},f_{D})=[0.60,0.76]$ with posterior content $0.85$.
This is similar to the results obtained using the unconditional prior but the
conditional prior puts more mass on larger values of the AUC hence the wider
plausible region with lower posterior content. Also, $c_{opt}(f_{ND},f_{D})=2$
with $Pl_{c_{opt}}(f_{ND},f_{D})=\{1,2\}$ with posterior probability content
approximately $1.00$ (actually $0.99999$)\ which reflects virtual certainty
that the true optimal value is in $\{1,2\}.$ $\blacksquare$

\subsection{Binormal diagnostic}

Suppose now that $X$ is a continuous diagnostic variable and it is assumed
that the distributions $F_{D}$ and $F_{ND}$ are normal distributions. The
assumption of normality should be checked by an appropriate test and it will
be assumed here that this has been carried out and normality was not rejected.
While the normality assumption may seem somewhat unrealistic, many aspects of
the analysis can be expressed in closed form and this allows for a deeper
understanding of ROC analyses more generally.

With $\Phi$ denoting the $N(0,1)$ cdf, then FNR$(c)=\Phi\left(  (c-\mu
_{D})/\sigma_{D}\right)  ,\ $FPR$(c)=1-\Phi\left(  (c-\mu_{ND})/\sigma
_{ND}\right)  $ so $c=\mu_{ND}+\sigma_{ND}\Phi^{-1}(1-(1-F_{ND}(c)))$ and%
\[
\text{AUC }=\int_{-\infty}^{\infty}\Phi\left(  \frac{\mu_{D}-\mu_{ND}}%
{\sigma_{D}}+\frac{\sigma_{ND}}{\sigma_{D}}z\right)  \varphi(z)\,dz.
\]
For given $(\mu_{D},\sigma_{D},\mu_{ND},\sigma_{ND})$ and $c,$ all these
values can be computed using $\Phi$ except the AUC and for that quadrature or
simulation via generating $z\sim N(0,1)$ is required.

The following results hold for the AUC with the proofs in the
Appendix.\smallskip\ 

\noindent\textbf{Lemma 2.} AUC $>1/2$ iff $\mu_{D}>\mu_{ND}$ and when $\mu
_{D}>\mu_{ND},$ the AUC is a strictly increasing function of $\sigma
_{ND}/\sigma_{D}.\smallskip$

\noindent From Lemma 2 it is clear that it makes sense to restrict the
parameterization so that $\mu_{D}>\mu_{ND}$ but we need to test the hypothesis
$H_{0}:\mu_{D}>\mu_{ND}$ first. Clearly Error$(c)=w$FNR$(c)+(1-w)$%
FPR$(c)\rightarrow1-w$ as $c\rightarrow-\infty$ and Error$(c)\rightarrow w$ as
$c\rightarrow\infty$ so, if Error$(c)$ does not achieve a minimum at a finite
value of $c,$ then the optimal cut-off is infinite and the optimal error is
$\min\{w,1-w\}.$ It is possible to give conditions under which a finite cutoff
exists and express $c_{opt}$ in closed form when the parameters and the
relevant prevalence $w$ are all known. $\smallskip$

\noindent\textbf{Lemma 3.} (i) When $\sigma_{D}^{2}=\sigma_{ND}^{2}=\sigma
^{2},$ then a finite optimal cut-off minimizing Error$(c)$ exists iff $\mu
_{D}>\mu_{ND}$ and in that case
\begin{equation}
c_{opt}=\frac{\mu_{D}+\mu_{ND}}{2}+\frac{\sigma^{2}}{\mu_{D}-\mu_{ND}}%
\log\left(  \frac{1-w}{w}\right)  . \label{bieq1}%
\end{equation}
(ii) When $\sigma_{D}^{2}\neq\sigma_{ND}^{2},$ then a finite optimal cut-off
exists iff
\begin{equation}
(\mu_{D}-\mu_{ND})^{2}+2\left(  \sigma_{D}^{2}-\sigma_{ND}^{2}\right)
\log\left(  \frac{1-w}{w}\frac{\sigma_{D}}{\sigma_{ND}}\right)  \geq0
\label{bieq2}%
\end{equation}
and in that case
\begin{equation}
c_{opt}=\frac{\sigma_{ND}^{2}\mu_{D}-\sigma_{D}^{2}\mu_{ND}}{\sigma_{ND}%
^{2}-\sigma_{D}^{2}}-\frac{\sigma_{ND}\sigma_{D}}{\sigma_{ND}^{2}-\sigma
_{D}^{2}}\left\{
\begin{array}
[c]{c}%
(\mu_{D}-\mu_{ND})^{2}+\\
2\left(  \sigma_{D}^{2}-\sigma_{ND}^{2}\right)  \log\left(  \frac{1-w}{w}%
\frac{\sigma_{D}}{\sigma_{ND}}\right)
\end{array}
\right\}  ^{1/2}. \label{bieq3}%
\end{equation}

\noindent Note that when $w=1/2,$ then in (i) $c_{opt}=(\mu_{D}+\mu_{ND})/2$
as one might expect. In the case of unequal variances there is an additional
restriction beyond $\mu_{D}\geq\mu_{ND}$ required to hold if the diagnostic is
to serve as a reasonable classifier. The following shows that these can be
combined in a natural way.$\smallskip$

\noindent\textbf{Corollary 4.} The restrictions $\mu_{D}\geq\mu_{ND}$ and
(\ref{bieq2}) hold iff%
\begin{equation}
\mu_{D}-\mu_{ND}-\left\{  \max\left[  0,-2\left(  \sigma_{D}^{2}-\sigma
_{ND}^{2}\right)  \log\left(  \frac{1-w}{w}\frac{\sigma_{D}}{\sigma_{ND}%
}\right)  \right]  \right\}  ^{1/2}\geq0. \label{bieq3a}%
\end{equation}
So, if one is unwilling to assume constant variance, then the hypothesis
$H_{0}:$ (\ref{bieq3a}) holds, needs to be assessed. There is some importance
to these results as they demonstrate that a finite optimal cutoff may in fact
not exist at least when considering both types of error. For example, when
$\mu_{ND}=1,\mu_{D}=2,\sigma_{D}=1,\sigma_{ND}=1.5,$ then for any
$w\leq0.30885,$ the optimal cutoff is $c_{opt}=\infty$ with Error$(\infty)=w.$
When $c_{opt}$ is infinite, then one may need to consider various cutoffs $c$
and find one that is acceptable at least with respect to\ some of the error
characteristics FNR$(c),$ FPR$(c),$ Error$(c)$, FDR$(c)$ and FNDR$(c).$

Consider now examples with equal and unequal variances.\smallskip

\noindent\textbf{Example 3. }\textit{Binormal with }$\sigma_{ND}^{2}%
=\sigma_{D}^{2}.$

There may be reasons why the assumption of equal variance is believed to hold
but this needs to be assessed and evidence in favor found. If evidence against
the assumption is found, then the approach of Example 4 can be used. A
possible prior is given by $\pi_{1}(\mu_{ND},\sigma^{2})\pi_{2}(\mu
_{D}\,|\,\sigma^{2})$ where
\[
\mu_{ND}\,|\,\sigma^{2}\sim N(\mu_{0},\tau_{0}^{2}\sigma^{2}),\text{ }\mu
_{D}\,|\,\sigma^{2}\sim N(\mu_{0},\tau_{0}^{2}\sigma^{2}),\text{ }1/\sigma
^{2}\sim\text{ gamma}(\lambda_{1},\lambda_{2})
\]
and this is a conjugate prior. The hyperparameters that need to be elicited
are $(\mu_{0},\tau_{0}^{2},\lambda_{1},\lambda_{2}).$ Consider first eliciting
the prior for $(\mu_{ND},\sigma^{2}).$ For this an interval $(m_{1},m_{2})$ is
specified such that is it believed that $\mu_{ND}\in(m_{1},m_{2})$ with
virtual certainty (say with probability $\gamma=0.99).$ Then putting $\mu
_{0}=(m_{1}+m_{2})/2$ implies
\[
\gamma\leq\Phi((m_{2}-\mu_{0})/\tau_{0}\sigma)-\Phi((m_{1}-\mu_{0})/\tau
_{0}\sigma)=2\Phi((m_{2}-m_{1})/2\tau_{0}\sigma)-1
\]
which implies $\sigma\leq(m_{2}-m_{1})/2\tau_{0}z_{(1+\gamma)/2}$ where
$z_{(1+\gamma)/2}=\Phi^{-1}((1+\gamma)/2).$ The interval $\mu_{ND}\pm\sigma
z_{(1+\gamma)/2}$ will contain an observation from $F_{ND\text{ }}$with
virtual certainty and let $(l_{0},u_{0})$ be lower and upper bounds on the
half-length of this interval so $l_{0}/z_{(1+\gamma)/2}\leq\sigma\leq
u_{0}/z_{(1+\gamma)/2}$ with virtual certainty. This implies $\tau_{0}%
=(m_{2}-m_{1})/2u_{0}.$ This leaves specifying the hyperparameters
$(\lambda_{1},\lambda_{2}),$\ and letting $G(\cdot,\lambda_{1},\lambda_{2})$
denote the cdf of the gamma$(\lambda_{1},\lambda_{2})$ distribution, then
$(\lambda_{1},\lambda_{2})$ satisfying
\begin{equation}
G(z_{(1+\gamma)/2}^{2}/l_{0}^{2},\lambda_{1},\lambda_{2})=(1+\gamma)/2,\text{
}G(z_{(1+\gamma)/2}^{2}/u_{0}^{2},\lambda_{1},\lambda_{2})=(1-\gamma)/2
\label{priorits1}%
\end{equation}
will give the specified $\gamma$ coverage. Noting that $G(x,\lambda
_{1},\lambda_{2})=G(\lambda_{2}x,\lambda_{1},1),$ first specify $\lambda_{1}$
and solve the first equation in (\ref{priorits1}) for $\lambda_{2}$ and then
solve the second equation in (\ref{priorits1}) for $\lambda_{1}$ and continue
this iteration until the values give a probability content to $(l_{0}%
/z_{(1+\gamma)/2},u_{0}/z_{(1+\gamma)/2})$ that is sufficiently close to
$\gamma$. 
Putting $s_{D}^{2}=||x_{D}-\bar{x}_{D}1||^{2},s_{ND}^{2}=||x_{ND}%
-\bar{x}_{ND}1||^{2},$ the posterior is then%
\begin{align*}
&  \mu_{ND}\,|\,\sigma^{2},x_{ND}\sim N\left(
\begin{array}
[c]{c}%
(n_{ND}+1/\tau_{0}^{2})^{-1}(n_{ND}\bar{x}_{ND}+\mu_{0}/\tau_{0}^{2}),\\
(n_{ND}+1/\tau_{0}^{2})^{-1}\sigma^{2}%
\end{array}
\right)  ,\\
&  \mu_{D}\,|\,\sigma^{2},x_{D}\sim N\left(
\begin{array}
[c]{c}%
(n_{D}+1/\tau_{0}^{2})^{-1}(n_{D}\bar{x}_{D}+\mu_{0}/\tau_{0}^{2}),\\
(n_{D}+1/\tau_{0}^{2})^{-1}\sigma^{2}%
\end{array}
\right)  ,\\
&  1/\sigma^{2}\,|\,(x_{ND},x_{D})\sim\text{ gamma}(\lambda_{1}+(n_{D}%
+n_{ND})/2,\lambda_{x})
\end{align*}
where $\lambda_{x}=\lambda_{2}+(s_{D}^{2}+s_{ND}^{2})/2+(n_{D}+1/\tau_{0}%
^{2})^{-1}(n_{D}/\tau_{0}^{2})(\bar{x}_{D}-\mu_{0})^{2}/2+(n_{ND}+1/\tau
_{0}^{2})^{-1}(n_{ND}/\tau_{0}^{2})(\bar{x}_{ND}-\mu_{0})^{2}/2.$

Suppose the following values of the mss were obtained based on samples of
$n_{ND}=25$ from $F_{ND}=N(0,1)$ and $n_{D}=20$ from $F_{D}=N(1,1)$%
\[
(\bar{x}_{ND},s_{ND}^{2})=(-0.072,19.638),\text{ }(\bar{x}_{D},s_{D}%
^{2})=(0.976,16.778).
\]
So the true values of the parameters are $\mu_{ND}=0,\mu_{D}=1,\sigma^{2}=1.$
In this case AUC $=\int_{-\infty}^{\infty}\Phi\left(  1+z\right)
\varphi(z)\,dz=0.760.$ Supposing that the relevant prevalence is
$w=0.4,c_{opt}=0.5+\log\left(  0.6/0.4\right)  =0.905,$ FNR$(c_{opt}%
)=\Phi\left(  0.905-1\right)  =0.46,\ $FPR$(c_{opt})=1-\Phi\left(
0.905\right)  =0.18,$ Error$(c_{opt})=0.30$,$\,$FDR$(c_{opt})=0.34,$
FNDR$(c_{opt})=0.27.$

For the prior elicitation, suppose it is known with virtual certainty that
both means lie in $(-5,5)$ and $(l_{0},u_{0})=(1,10)$ so we take $\mu
_{0}=(-5+5)/2=0,\tau_{0}=(m_{2}-m_{1})/2u_{0}=0.5$ and the iterative process
leads to $(\lambda_{1},\lambda_{2})=(1.787,1.056).$ For inference about
$c_{opt}$ it is necessary to specify a prior distribution for the prevalence
$w.$ This can range from $w$ being completely known to being completely
unknown whence a uniform(0,1) (beta$(1,1)$) would be appropriate. Following
the developments of Section 3.1, suppose it is known that $w\in\lbrack
l,u]=[0.2,0.6]$ with prior probability $\gamma=0.99,$ so in this case $\xi
_{w}=(l+u)/2=0.4$ and $\tau_{w}=35.89725$ and the prior is $w\sim$
beta$(15.3589,22.53835).$

The first inference step is to assess the hypothesis $H_{0}:$ AUC $>1/2$ which
is equivalent to $H_{0}:$ $\mu_{ND}<\mu_{D}$ by computing the prior and
posterior probabilities of this event to obtain the relative belief ratio. The
prior probability of $H_{0}$ given $\sigma^{2}$ is $\int_{-\infty}^{\infty
}\Phi\left(  (\mu_{D}-\mu_{0})/\tau_{0}\sigma\right)  (\tau_{0}\sigma
)^{-1}\varphi\left(  (\mu_{D}-\mu_{0})/\tau_{0}\sigma\right)  \,d\mu_{D}=1/2$
and averaging this quantity over the prior for $\sigma^{2}$ we get $1/2.$ The
posterior probability of this event can be easily obtained via simulating from
the joint posterior. When this is done in the specific numerical example, the
relative belief ratio of this event is $2.011$ with posterior content $0.999$
so there is strong evidence that $H_{0}:$ AUC $>1/2$ is true.

If evidence is found against $H_{0},$ then this would indicate a poor
diagnostic. If evidence is found in favor, then we can proceed conditionally
given that $H_{0}$ holds and so condition the joint prior and joint posterior
on this event being true when making inferences about AUC, $c_{opt},$ etc. So
for the prior it is necessary to generate $1/\sigma^{2}\sim$ gamma$(\alpha
_{0},\beta_{0})$ and then generate $(\mu_{D},\mu_{ND})$ from the joint
conditional prior given $\sigma^{2}$ and that $\mu_{D}>\mu_{ND}.$ Denoting the
conditional priors given $\sigma^{2}$ by $\pi_{D}(\mu_{D}\,|\,\sigma^{2})$ and
$\pi_{ND}(\mu_{ND}\,|\,\sigma^{2}),$ we see that this joint conditional prior
is proportional to
\begin{align*}
&  \pi_{ND}(\mu_{ND}\,|\,\sigma^{2})\pi_{D}(\mu_{D}\,|\,\sigma^{2})\\
&  =\Pi_{ND}(\mu_{ND}<\mu_{D}\,|\,\mu_{D},\sigma^{2})\frac{\pi_{ND}(\mu_{ND}%
)}{\Pi_{ND}(\mu_{ND}<\mu_{D}\,|\,\sigma^{2})}\pi_{D}(\mu_{D}\,|\,\sigma^{2}).
\end{align*}
While generally it is not possible to generate efficiently from this
distribution we can use importance sampling to calculate any expectations by
generating $\mu_{D}\sim\mu_{D}\,|\,\sigma^{2}\sim N(\mu_{0},\tau_{0}^{2}%
\sigma^{2}),\mu_{ND}\sim N(\mu_{0},\tau_{0}^{2}\sigma^{2}\,|\,(-\infty,\mu
_{D}])$ with $\Pi_{ND}(\mu_{ND}<\mu_{D}\,|\,\mu_{D},\sigma^{2})=\Phi((\mu
_{D}-\mu_{0})/\tau_{0}\sigma)$ serving as the importance sampling weight and
where $N(\mu_{0},\tau_{0}^{2}\sigma^{2}\,|\,(-\infty,\mu_{D}])$ denotes the
$N(\mu_{0},\tau_{0}^{2}\sigma^{2})$ distribution conditioned to $(-\infty
,\mu_{D}]$ with density $\Phi^{-1}\left(  (\mu_{D}-\mu_{0})/\tau_{0}%
\sigma\right)  (2\pi\tau_{0}^{2}\sigma^{2})^{-1/2}\varphi\left(  (\mu_{ND}%
-\mu_{0})/\tau_{0}\sigma\right)  $ for $\mu_{ND}\leq\mu_{D}$ and 0 otherwise.
Generating from this distribution via inversion is easy since the cdf is
$\Phi\left(  (\mu_{ND}-\mu_{0})/\tau_{0}\sigma\right)  /\Phi\left(  (\mu
_{D}-\mu_{0})/\tau_{0}\sigma\right)  .$ Note that, if we take the posterior
from the unconditioned prior and condition that, we will get the same
conditioned posterior as when we use the conditioned prior to obtain the
posterior. This implies that in the joint posterior for $(\mu_{ND},\mu
_{D},\sigma^{2})$ it is only necessary to adjust the posterior for $\mu_{ND}$
as was done with the prior and this is also easy to generate from. Note that
Lemma 3(i) implies that it is necessary to use the conditional prior and
posterior to guarantee that $c_{opt}$ exists finitely.

\begin{figure}
[t]
\begin{center}
\includegraphics[width=8cm]%
{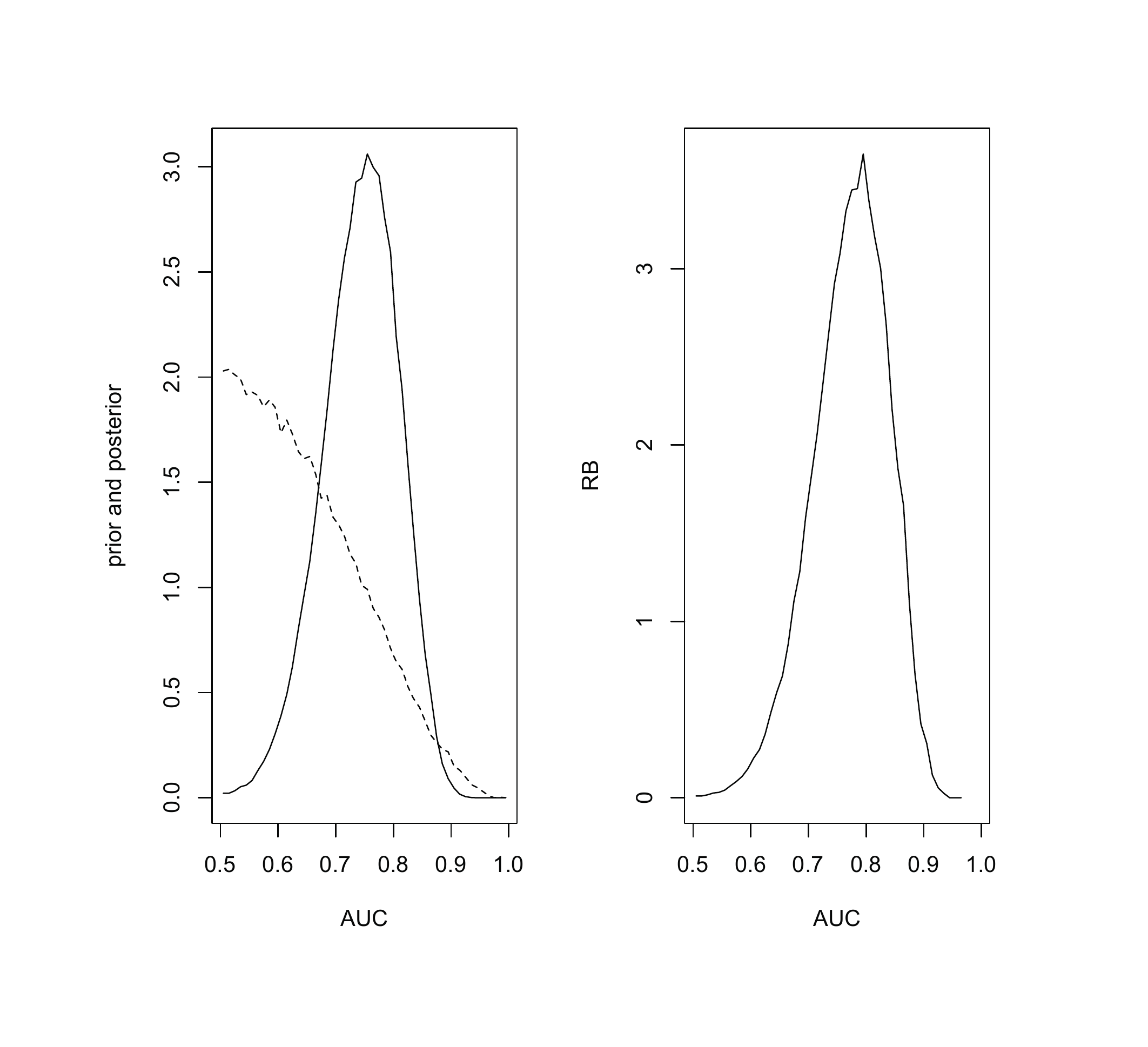}%
\caption{The conditioned prior (- -) and posterior (--) densities (left panel)
and the relative belief ratio (right panel) of the AUC in Example 3.}%
\label{binorm1}%
\end{center}
\end{figure}

Since $H_{0}$ was accepted, the conditional sampling was implemented and the
estimate of the AUC is $0.795$ with plausible region $[0.670,0.880]$ which has
posterior content $0.856.$ So the estimate is close to the true value but
there is substantial uncertainty. Figure \ref{binorm1} is a plot of the
conditioned prior, the conditioned posterior and relative belief ratio for
this data.

\begin{figure}
[t]
\begin{center}
\includegraphics[width=7cm]%
{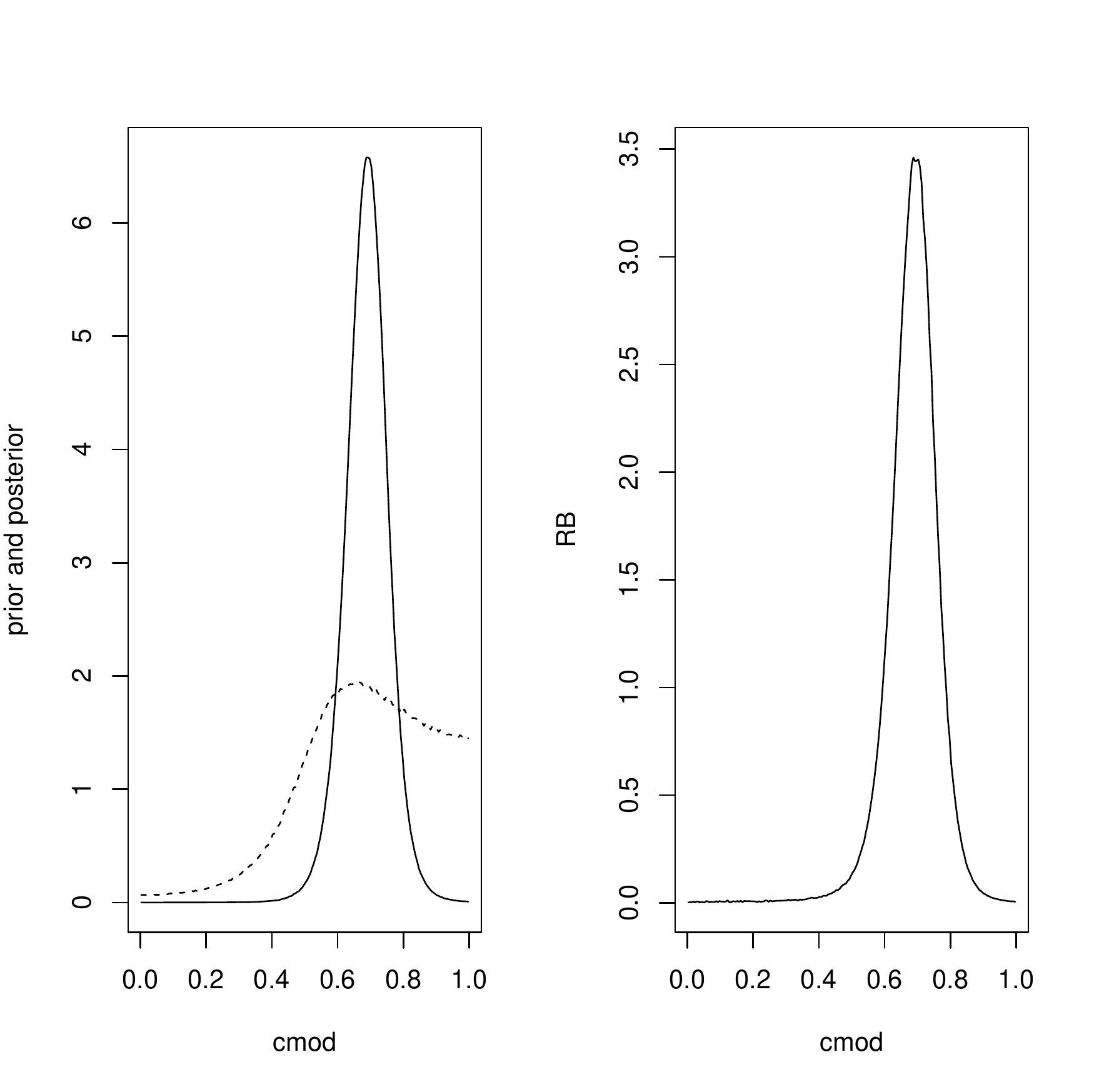}%
\caption{Plots of the prior (- -), posterior (left panel) and relative belief
ratio (right panel) of $c_{opt}$ in Example 3.}%
\label{binorm2}%
\end{center}
\end{figure}

With the specified prior for $w,$ the posterior based on the given data is
beta$(35.3589,47.53835)$ which leads to estimate $0.444$ for $w$ with
plausible interval $(0.374,0.516)$ having posterior probability content
$0.782.$ Using this prior and posterior for $w$ and the conditioned prior and
posterior for $(\mu_{D},\mu_{ND},\sigma^{2}),$ we proceed to inference about
$c_{opt}$ and the error characteristics associated with this classification. A
computational problem arises when obtaining the prior and posterior
distributions of $c_{opt}$ as it is clear from (\ref{bieq1}) that these
distributions can be extremely long-tailed. As such, we transform to
$c_{\operatorname{mod}}=0.5+\arctan(c_{opt})/\pi\in\lbrack0,1]$ (the Cauchy
cdf), obtain the estimate $c_{\operatorname{mod}}(d)$ where $d=(n_{ND},\bar
{x}_{ND},s_{ND}^{2},n_{D},\bar{x}_{D},s_{D}^{2})$ and its plausible region and
then, applying the inverse transform, obtain $c_{opt}(d)=\tan(\pi
(c_{\operatorname{mod}}(d)-0.5))$ and its plausible region. It is notable that
relative belief inferences are invariant under 1-1 smooth transformations, so
it doesn't matter which parameterization is used, but it is much easier
computationally to work with a bounded quantity. Also, if a shorter tailed cdf
is used rather than a Cauchy, e.g. a $N(0,1)$ cdf, then errors can arise due
to extreme negative values being always transformed to 0 and very extreme
positive values always transformed to 1. Figure \ref{binorm2} is a plot of the
prior density, posterior density and relative belief ratio of
$c_{\operatorname{mod}}.$ For this data $c_{opt}(d)=0.715$ with plausible
interval $(0.316,1.228)$ having posterior content $0.860.$ Large Monte Carlo
samples were used to get smooth estimates of the densities and relative belief
ratio but these only required a few minutes of computer time on a desktop. The
estimated error characteristics at this value of $c_{opt}$ are as follows:
FNR$(0.715)=0.41$, FPR$(0.715)=0.22$, Error$(0.715)=0.27$, FDR$(0.715)=0.30$,
FNDR$(0.715)=0.24$ which are close to the true values. $\blacksquare
\smallskip$%

\noindent\textbf{Example 4. }\textit{Binormal with }$\sigma_{ND}^{2}\neq
\sigma_{D}^{2}.$

In this case the prior is given by $\pi_{1}(\mu_{ND},\sigma_{ND}^{2})\pi
_{2}(\mu_{D},\sigma_{D}^{2})$ where
\begin{align}
&  \mu_{ND}\,|\,\sigma_{ND}^{2}\sim N(\mu_{0},\tau_{0}^{2}\sigma_{ND}%
^{2}),\text{ }1/\sigma_{ND}^{2}\sim\text{ gamma}(\lambda_{1},\lambda
_{2})\nonumber\\
&  \mu_{D}\,|\,\sigma_{D}^{2}\sim N(\mu_{0},\tau_{0}^{2}\sigma_{D}^{2}),\text{
}1/\sigma_{D}^{2}\sim\text{ gamma}(\lambda_{1},\lambda_{2}). \label{priorex4}%
\end{align}
Although this specifies the same prior for the two populations, this is easily
modified to use different priors and, in any case, the posteriors are
different. Again it is necessary to check that the AUC $>1/2$ but also to
check that $c_{opt}$ exists finitely using the full posterior based on this prior and
for this we have the hypothesis $H_{0}$ given by Corollary 4. If evidence in
favor of $H_{0}$ is found, the prior is replaced by the conditional prior
given this event for inference about $c_{opt}.$ This can be implemented via
importance sampling as was done in Example 3 and similarly for the posterior.

\begin{figure}
[t]
\begin{center}
\includegraphics[width=8cm]%
{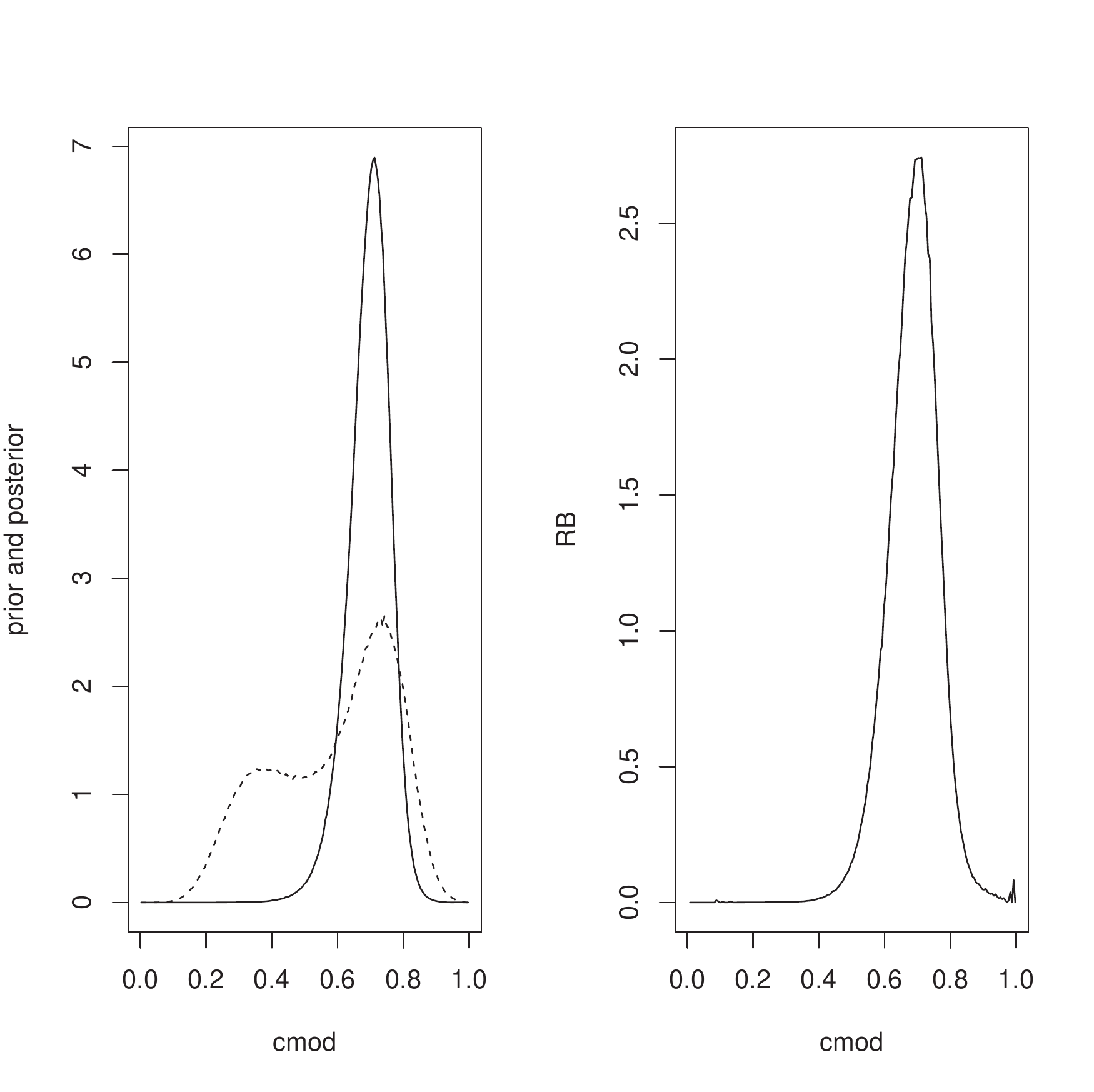}%
\caption{Plots of the prior (- -), posterior (left panel) and relative belief
ratio (right panel) of $c_{opt}$ in Example 4.}%
\label{binorm3}%
\end{center}
\end{figure}

Using the same data and hyperparameters as in Example 3 the relative belief
ratio of $H_{0}$ is $3.748$ with posterior content $0.828$ so there is
reasonably strong evidence in favor of $H_{0}.$ Estimating the value of the
AUC is then based on conditioning on $H_{0}$ being true. Using the conditional
prior given that $H_{0}$ is true, the relative belief estimate of the AUC is
$0.793$ with plausible interval $(0.683,0.857)$ with posterior content
$0.839.$ The optimal cutoff is estimated as $c_{opt}(d)=0.739$ with plausible
interval $(0.316,1.228)$ having posterior content $0.875.$ Figure
\ref{binorm3} is a plot of the prior density, posterior density and relative
belief ratio of $c_{\operatorname{mod}}.$ The estimates of the error
characteristics at $c_{opt}(d)$ are as follows: FNR$(0.739)=0.43$,
FPR$(0.739)=0.19$, Error$(0.739)=0.28$, FDR$(0.739)=0.28$,
FNDR$(0.624)=0.264.$

It is notable that these inferences are very similar to those in Example 3. It
is also noted that the sample sizes are not big and so the only situation
where it might be expected that the inferences will be quite different between
the two analyses is when the variances are substantially different.
$\blacksquare$%

\subsection{Nonparametric Bayes model}

Suppose that $X$ is a continuous variable, of course still measured to some
finite accuracy, and available information is such that no particular finite
dimensional family of distributions is considered feasible. The situation is
considered where a normal distribution $N(\mu,\sigma^{2}),$ perhaps after
transforming the data, is considered as a possible base distribution for $X$
but we want to allow for deviation from this form. Alternative choices can
also be made for the base distribution. The statistical model is then to
assume that the $x_{ND}$ and $x_{D}$ are generated as samples from $F_{ND}$
and $F_{D},$ where these are independent values from a DP$(a,H)$ (Dirichlet)
process with base $H=N(\mu,\sigma^{2})$ for some $(\mu,\sigma^{2})\ $and
concentration parameter $a.$ Actually, since it is difficult to argue for some
particular choice of $(\mu,\sigma^{2}),$ it is supposed that $(\mu,\sigma
^{2})$ is generated from\ a prior $\pi(\mu,\sigma^{2}).$ The prior on
$(F_{ND},F_{D})$ is then specified hierarchically as a mixture Dirichlet
process,
\begin{align*}
&  (\mu_{ND},\sigma_{ND}^{2})\sim\pi\text{ independent of }\,(\mu_{D}%
,\sigma_{D}^{2})\sim\pi,\\
&  F_{ND}\,|\,(\mu_{ND},\sigma_{ND}^{2})\sim\text{ DP}(a_{ND},N(\mu
_{ND},\sigma_{ND}^{2}))\text{\ independent of }\\
&  F_{D}\,|\,(\mu_{D},\sigma_{D}^{2})\sim\text{ DP}(a_{D},N(\mu_{D},\sigma
_{D}^{2}))\text{.}%
\end{align*}
To complete the prior it is necessary to specify $\pi$ and the concentration
parameters $a_{ND}$ and $a_{D}.$ For $\pi$ the prior is taken to be a normal
distribution elicited as discussed in Section 3.3 although other choices are
possible. For eliciting the concentration parameters, consider how strongly it
is believed that normality holds and for convenience suppose $a=a_{ND}=a_{D}.$
If $F\sim$ DP$(a,H)$ with $H$ a probability measure, then $E(F(A))=H(A)$ and
$Var(F(A))=H(A)(1-H(A))/(1+a).$ When $F$ a random measure from $P,$ then
$\sup_{A}P(|F(A)-H(A)|\geq\varepsilon)=\sup_{A}\{1-P(\max(0,H(A)-\varepsilon
)<F(A)<\min(1,H(A)+\varepsilon))\}$ which, when $P=$ DP$(a,H),$ equals
\begin{equation}
\sup_{r\in\lbrack0,1]}\{1-B([\max(0,r-\varepsilon),\min(1,r+\varepsilon
)],ar,a(1-r))\} \label{bound}%
\end{equation}
where $B(\cdot,\beta_{1},\beta_{2})$ denotes the beta$(\beta_{1},\beta_{2})$
measure. This upper bound on the probability that the random $F$ differs from
$H$ by at least $\varepsilon$ on an event can be made as small as desirable by
choosing $a$ large enough. For example, if $\varepsilon=0.25$ and it is
required that this upper bound be less than $0.1,$ then this satisfied when
$a\geq9.8$ and if instead $\varepsilon=0.1$, then $a\geq66.8$ is necessary.
Note that, since this bound holds for every continuous probability measure
$H,$ it also holds when $H$ is random, as considered here. So $a$ is
controlling how close it is believed that the true distribution is to $H$.
Alternative methods for eliciting $a$ can be found in Swartz(1993, 1999).

Generating $(F_{ND},F_{D})$ from the prior for given $(a,H)$ can only be done
approximately and the approach of Ishwaran and Zarepour (2002) is adopted. For
this, integer $n^{\ast}$ is specified and the measure $P_{n^{\ast}}=\sum
_{i=1}^{n^{\ast}}p_{i,n^{\ast}}I_{\{c_{i}\}}$ is generated where
$(p_{1,n^{\ast}},\ldots,p_{n^{\ast},n^{\ast}})\sim$ Dirichlet$(a/n^{\ast
},\ldots,.a/n^{\ast})$ independent of $c_{1},\ldots,c_{n^{\ast}}\overset
{iid}{\sim}H,$ since $P_{n^{\ast}}\overset{w}{\rightarrow}$ DP$(a,H)$ as
$n^{\ast}\rightarrow\infty.$ So to carry out a priori calculations proceed as
follows. Generate
\begin{align*}
&  (p_{ND1,n^{\ast}},\ldots,p_{NDn^{\ast},n^{\ast}})\sim\text{Dirichlet}%
((a/n^{\ast})\mathbf{1}_{n^{\ast}}),\text{ }(\mu_{ND},\sigma_{ND}^{2})\sim
\pi,\\
&  (c_{ND1},\ldots,c_{NDn^{\ast}})\,|\,(\mu_{ND},\sigma_{ND}^{2}%
)\overset{i.i.d.}{\sim}N(\mu_{ND},\sigma_{ND}^{2}),\text{ }w\sim
\text{beta}(\alpha_{1w},\alpha_{2w})
\end{align*}
and similarly for $(p_{D1,n^{\ast}},\ldots,p_{Dn^{\ast},n^{\ast}}),(\mu
_{D},\sigma_{D}^{2}),$ and $(c_{D1},\ldots,c_{Dn^{\ast}}).$ Then
\newline$F_{ND,n^{\ast}}(c)=\sum_{\{i:c_{NDi}\leq c\}}p_{NDin^{\ast}}$ is the
random cdf at $c\in R^{1}$ and similarly for $F_{D,n^{\ast}}$ and
AUC$=\sum_{i=1}^{n^{\ast}}(1-F_{D,n^{\ast}}(c_{NDi}))p_{NDi,n^{\ast}}\ $is a
value from the prior distribution of the AUC. This is done repeatedly to get
the prior distribution of the AUC as in our previous discussions and we
proceed similarly for the other quantities of interest.

The posterior given $(\mu_{ND},\sigma_{ND}^{2},\mu_{D},\sigma_{D}^{2})$ is
\begin{align*}
&  F_{ND}\,|\,x_{ND},(\mu_{ND},\sigma_{ND}^{2})\sim DP(a+n_{ND},H_{ND})\text{
independent of }\\
&  F_{D}\,|\,x_{D},(\mu_{D},\sigma_{D}^{2})\sim DP(a+n_{D},H_{D})
\end{align*}
with $H_{ND}(c)=a\Phi((c-\mu_{ND})/\sigma_{ND})/(a+n_{ND})+n_{ND}\hat{F}%
_{ND}(c)/(a+n_{ND})$ and $\hat{F}_{ND}(c)=\sum_{i=1}^{n_{ND}}I_{(-\infty
,c]}(x_{NDi})/n_{ND}$ is the empirical cdf (ecdf) based on $x_{ND}$ and
similarly for $H_{D}.$ The posteriors of $(\mu_{ND},\sigma_{ND}^{2})$ and
$(\mu_{D},\sigma_{D}^{2})$ are obtained via results in Antoniak (1974) and
Doss (1994). The posterior density of $(\mu_{ND},\sigma_{ND}^{2})$ given
$x_{ND}$ is proportional to
\[
\pi(\mu_{ND},\sigma_{ND}^{2})\prod\nolimits_{i=1}^{\tilde{n}_{ND}}\sigma
_{ND}^{-1}\varphi((\tilde{x}_{NDi}-\mu_{ND})/\mu_{ND})
\]
where $\tilde{n}_{ND}$ is the number of unique values in $x_{ND}$ and
$\{\tilde{x}_{ND1},\ldots,\tilde{x}_{ND\tilde{n}_{ND}}\}$ is the set of unique
values with mean $\tilde{x}_{ND}$ and sum of squared deviations $\tilde
{s}_{ND}^{2}$. From this it is immediate that
\begin{align*}
&  \mu_{ND}\,|\,\sigma_{ND}^{2},x_{ND}\sim N\left(
\begin{array}
[c]{c}%
(\tilde{n}_{ND}+1/\tau_{0}^{2})^{-1}(\tilde{n}_{ND}\tilde{x}_{ND}+\mu_{0}%
/\tau_{0}^{2}),\\
(\tilde{n}_{NDND}+1/\tau_{0}^{2})^{-1}\sigma_{ND}^{2}%
\end{array}
\right)  ,\\
&  1/\sigma_{ND}^{2}\,|\,x_{ND}\sim\text{ gamma}(\alpha_{0}+\tilde{n}%
_{ND}/2,\tilde{\lambda}_{x_{ND}})
\end{align*}
where $\tilde{\lambda}_{x_{ND}}=\lambda_{0}+\tilde{s}_{ND}^{2}/2+(\tilde
{n}_{ND}+1/\tau_{0}^{2})^{-1}(\tilde{n}_{ND}/\tau_{0}^{2})(\tilde{x}_{ND}%
-\mu_{0})^{2}/2.$ A similar result holds for the posterior of $(\mu_{D}%
,\sigma_{D}^{2}).$

To approximately generate from the full posterior specify some $n^{\ast\ast},$
put $p_{a,n_{ND}}=a/(a+n_{ND}),q_{a,n_{ND}}=1-p_{a,n_{ND}}$ and generate
\begin{align*}
&  (p_{ND1,n^{\ast\ast}},\ldots,p_{NDn^{\ast\ast},n^{\ast\ast}})\,|\,x_{ND}%
\sim\,\text{Dirichlet}(((a+n_{ND})/n^{\ast\ast})\mathbf{1}_{n^{\ast\ast}}),\\
&  (\mu_{ND},\sigma_{ND}^{2})\,|\,x_{ND}\sim\pi\left(  \cdot\,|\,x_{ND}%
\right)  ,\\
&  (c_{ND1},\ldots,c_{NDn^{\ast\ast}})\,|\,(\mu_{ND},\sigma_{ND}^{2}%
),x_{ND}\overset{i.i.d.}{\sim}p_{a,n_{ND}}N(\mu_{ND},\sigma_{ND}%
^{2})+q_{a,n_{ND}}\hat{F}_{ND},\\
&  w\,|\,x_{ND}\sim\,\text{beta}(\alpha_{1w}+n_{D},\alpha_{2w}+n_{ND})
\end{align*}
and similarly for $(p_{D1,n^{\ast\ast}},\ldots,p_{Dn^{\ast\ast},n^{\ast\ast}%
}),(\mu_{D},\sigma_{D}^{2})$ and $(c_{D1},\ldots,c_{Dn^{\ast\ast}}).$ If the
data does not comprise a sample from the full population, then the posterior
for $w$ is replaced by its prior.

There is an issue that arises when making inference about $c_{opt},$ namely,
the distributions for $c_{opt}$ that arises from this approach can be very
irregular and particularly the posterior distribution. In part this is due to
the discreteness of the posterior distributions of $F_{ND}$ and $F_{D}$ . This
doesn't affect the prior distribution because the points on which the
generated distributions are concentrated vary quite continuously among the
realizations and this leads to a relatively smooth prior density for
$c_{opt}.$ For the posterior, however, the sampling from the ecdf leads to a
very irregular, multimodal density for $c_{opt}.$ So some smoothing is
necessary in this case.

Consider now applying such an analysis to the dataset of Example 3, where we
know the true values of the quantities of interest and then to a dataset
concerned with the COVID-19 epidemic.$\smallskip$

\noindent\textbf{Example 5. }\textit{Binormal data (Examples 3 and 4)}

The data used in Example 3 is now analyzed but using the methods of this
section. The prior on $(\mu_{ND},\sigma_{ND}^{2}),(\mu_{D},\sigma_{D}^{2})$
and $w$ is taken to be the same as that used in Example 4 so the variances are
not assumed to be the same. The value $\varepsilon=0.25$ is used and requiring
(\ref{bound}) to be less than $0.018$ leads to $a=20.$ So the true
distributions are allowed to differ quite substantially from a normal
distribution. Testing the hypothesis $H_{0}:$ AUC $>1/2$ led to the relative
belief ratio $1.992$ (maximum possible value is $2$) and the strength of the
evidence is $0.997$ so there is strong evidence that $H_{0}$ is true. The AUC,
based on the prior conditioned on $H_{0}$ being true, is estimated to be equal
to $0.839$ with plausible interval $(0.691,0.929)$ having posterior content
$0.814.$ For this data $c_{opt}(d)=0.850$ with plausible interval
$(0.45,1.75)$ having posterior content $0.835$. The true value of the AUC is
$0.760$ and the true value of $c_{opt}$ is $0.905$ so these inferences are
certainly reasonable although, as one might expect, when the length of the
plausible intervals are taken into account, they are not as accurate as those
when binormality is assumed as this is correct for this data. So the DP
approach worked here although the posterior density for $c_{opt}$ was quite
multimodal and required some smoothing (averaging 3 consecutive values).
$\blacksquare\smallskip$

\noindent\textbf{Example 6. }\textit{COVID-19 data}

A dataset was downloaded from https://github.com/YasinKhc/Covid-19 containing
data on 3397 individuals diagnosed with COVID-19 and includes whether or not
the patient survived the disease, their gender and their age. There are 1136
complete cases on these variables of which 646 are male, with 52 having died,
and 490 are female, with 25 having died. Our interest is in the use of a
patient's age $X$ to predict whether or not they will survive. More detail on
this dataset can be found in Charvadeh and Yi (2020). The goal is to determine
a cutoff age so that extra medical attention can be paid to patients beyond
that age. Also it is desirable to see whether or not gender leads to
differences so separate analyses can be carried out by gender. So, for
example, in the male group ND refers to those males with COVID-19 that will
not die and D refers to the population that will. Looking at histograms of the
data, it is quite clear that binormality is not a suitable assumption and no
transformation of the age variable seems to be available to make a normality
assumption more suitable. Table \ref{tabex5} gives summary statistics for the
subgroups. Of some note is that condition\ (\ref{bieq3a}), when using
standard\ estimates for population quantities like $w=52/646=0.08$ for
Males\ and $w=25/490=0.05$ for females, is not satisfied which suggests that
in a binormal analysis no finite optimal cutoff exists.%

\begin{table}[tbp] \centering
\begin{tabular}
[c]{|l|l|l|l|l|l|}\hline
Group & number & mean & std. dev. & min & max\\\hline
ND males & \multicolumn{1}{|r|}{$594$} & \multicolumn{1}{|r|}{$48.81$} &
\multicolumn{1}{|r|}{$17.72$} & \multicolumn{1}{|r|}{$0.50$} &
\multicolumn{1}{|r|}{$85.00$}\\
D males & \multicolumn{1}{|r|}{$52$} & \multicolumn{1}{|r|}{$68.46$} &
\multicolumn{1}{|r|}{$13.66$} & \multicolumn{1}{|r|}{$36.00$} &
\multicolumn{1}{|r|}{$89.00$}\\
ND females & \multicolumn{1}{|r|}{$465$} & \multicolumn{1}{|r|}{$48.69$} &
\multicolumn{1}{|r|}{$18.73$} & \multicolumn{1}{|r|}{$2.00$} &
\multicolumn{1}{|r|}{$96.00$}\\
D females & \multicolumn{1}{|r|}{$25$} & \multicolumn{1}{|r|}{$77.36$} &
\multicolumn{1}{|r|}{$12.12$} & \multicolumn{1}{|r|}{$48.00$} &
\multicolumn{1}{|r|}{$95.00$}\\\hline
\end{tabular}
\caption{ Summary statistics for the data in Example 6.}\label{tabex5}%
\end{table}%

For the prior, it is assumed that $(\mu_{ND},\sigma_{ND}^{2})$ and $(\mu
_{D},\sigma_{D}^{2})$ are independent values from the same prior distribution
as in (\ref{priorex4}). For the prior elicitation suppose it is known with
virtual certainty that both means lie in $(20,70)$ and $(l_{0},u_{0})=(20,50)$
so we take $\mu_{0}=45,\tau_{0}=(m_{2}-m_{1})/2u_{0}=0.75$ and the iterative
process leads to $(\lambda_{1},\lambda_{2})=(8.545,1080.596)$ which implies a
prior on the $\sigma$'s with mode at $10.932$ and the interval
$(7.764,19.411)$ containing $0.99$ of the prior probability$.$ Here the
relevant prevalence refers to the proportion of COVID-19 patients that will
die and it is supposed that $w\in\lbrack0.00,0.15]$ with virtual certainty
which implies $w\sim$ beta$(9.81,109.66).$ So the prior probability that
someone with COVID-19 will die is assumed to be less than 15\% with virtual
certainty. Since normality is not an appropriate assumption for the
distribution of $X,$ the choice $\varepsilon=0.25$ with the upper bound
(\ref{bound}) equal to $0.1$ seems reasonable and so $a=9.8.$ This specifies
the prior that is used for the analysis with both genders and it is to be
noted that it is not highly informative.

For males the hypothesis AUC $>1/2$ is assessed and $RB=1.991$ (maximum value
2) with strength effectively equal to $1.00$ was obtained, so there is
extremely strong evidence that this is true. The unconditional estimate of the
AUC is $0.808$ with plausible region $[0.698,0.888]$ having posterior content
$0.959,$ so there is a fair bit of uncertainty concerning the true value. For
the conditional analysis, given that AUC $>1/2,$ the estimate of the AUC is
$0.806$ with plausible region $[0.731,0.861]$ having posterior content
$0.932.$ So the conditional analysis gives a similar estimate for the AUC with
a small increase in accuracy.\ In either case it seems that the AUC is
indicating that Age should be a reasonable diagnostic. Note that the standard
nonparametric estimate of the AUC is $0.810$ so the two approaches agree here.
For females the hypothesis AUC $>1/2$ is assessed and $RB=1.994$ with strength
effectively equal to $1$ was obtained, so there is extremely strong evidence
that this is true. The unconditional estimate of the AUC is $0.873$ with
plausible region $(0.742,0.948)$ having posterior content $0.968$. For the
conditional analysis, given that AUC $>1/2,$ the estimate of the AUC is
$0.874$ with plausible region $(0.791,0.936)$ having posterior content
$0.956.$ The traditional estimate of the AUC is $0.902$ so the two approaches
are again in close agreement.

Inferences for $c_{opt}$ are more problematical in both genders. Consider the
male data. The data set is very discrete as there are many repeats and the
approach samples from the ecdf about 84\% of the time for the males that died
and 98\% of the time for the males that didn't die. The result is a plausible
region that is not contiguous even with smoothing. Without smoothing the
estimate is $c_{opt}(d)=85.5$ for males$,$ which is a very dominant peak for
the relative belief ratio. The plausible region contains $0.928$ of the
posterior probability and, although it is not a contiguous interval, the
subinterval $[85.2,85.8]$ is a $0.58$-credible interval for $c_{opt}$ that is
in agreement with the evidence. If we continuize the data by adding a
uniform(0,1) random error to each age in the data set, then $c_{opt}(d)=86.1$
and plausible interval $[75.9,86.7]$ with posterior content $0.968$ is
obtained. These cutoffs are both greater than the maximum value in the ND
data, so there is ample protection against false positives but it is
undoubtedly false negatives that are of most concern in this context. If
instead the FNDR is used as the error criterion to minimize, then
$c_{opt}(d)=35.7$ and plausible interval $[26.1,35.7]$ with posterior content
$0.826$ is obtained and so in this case there will be too many false
positives. So a useful optimal cutoff incorporating the relevant prevalence
does not seem to exist with this data.

If the relevant prevalence is ignored and $w_{0}$FNR$+(1-w_{0})$FPR is used
for some fixed weight $w_{0}$ to determine $c_{opt}(d)$, \ then more
reasonable values are obtained. Table \ref{maletab} gives the estimates for
various $w_{0}$ values. With $w_{0}=0.5$ (corresponding to using Youden's
index) $c_{opt}(d)=65.7$ while if $w_{0}=0.7,$ then $c_{opt}(d)=56.7.$ When
$w_{0}$ is too small or too large then the value of $c_{opt}(d)$ is not
useful. While these estimates do not depend on the relevant prevalence, the
error characteristics that do depend on this prevalence (as expressed via its
prior and posterior distributions) can still be quoted and a decision made as
to whether or not to use the diagnostic. Table \ref{maletab2} contains the
estimates of the error characteristics at $c_{opt}(d)$ for various values of
$w_{0}$ where these are determined using the prior and posterior on the
relevant prevalence $w.$ Note that these estimates are determined as the
values that maximize the corresponding relative belief ratios and take into
account the posterior of $w.$ So, for example, the estimate of the Error is
not the convex combination of the estimates of FNR and FPR\ based on the
$w_{0}$ weight. Another approach is to simply set the cutoff Age at a value at
a value $c_{0}$ and then investigate the error characteristics at that value.
For example, with $c_{0}=60,$ then the estimated values are given by
FNR$(c_{0})=0.238,$ FPR$(c_{0})=0.308,$ Error$(c_{0})=0.328,$ FDR$(c_{0})=$
$0.818$ and FNDR$(c_{0})=0.028.$

Similar results are obtained for the cutoff with female data although with
different values. Overall, Age by itself does not seem to be useful classifier
although that is a decision for medical practitioners. Perhaps it is more
important to treat those who stand a significant chance of dying more
extensively and not worry too much that some treatments are not necessary. The
clear message from this data, however, is that a relatively high AUC does not
immediately imply that a diagnostic is useful and the relevant prevalence is a
key aspect of this determination.%

\begin{table}[tbp] \centering
\begin{tabular}
[c]{|l|l|l|}\hline
$w_{0}=$ weight of FNR & $c_{opt}(d)$ & plausible range (post. prob.)\\\hline
\multicolumn{1}{|c|}{$0.1$} & \multicolumn{1}{|c|}{$85.5$} &
\multicolumn{1}{|c|}{$75.3-118.5$ $(0.945)$}\\
\multicolumn{1}{|c|}{$0.3$} & \multicolumn{1}{|c|}{$65.1$} &
\multicolumn{1}{|c|}{$64.5-85.5$ $(0.868)$}\\
\multicolumn{1}{|c|}{$0.5$} & \multicolumn{1}{|c|}{$65.1$} &
\multicolumn{1}{|c|}{$55.5-72.3$ $(0.939)$}\\
\multicolumn{1}{|c|}{$0.7$} & \multicolumn{1}{|c|}{$56.7$} &
\multicolumn{1}{|c|}{$35.7-58.5$ $(0.919)$}\\
\multicolumn{1}{|c|}{$0.9$} & \multicolumn{1}{|c|}{$35.7$} &
\multicolumn{1}{|c|}{$33.3-52.5$ $(0.875)$}\\\hline
\end{tabular}
\caption{Weighted error $w_0$FNR+($1-w_0)$FPR determining $c_{opt} (d)$ for Males in Example 6.}\label{maletab}%
\end{table}%
%

\begin{table}[tbp] \centering
\begin{tabular}
[c]{|l|l|l|l|l|l|}\hline
$w_{0}=$ weight of FNR & FNR & FPR & Error & FDR & FNDR\\\hline
\multicolumn{1}{|c|}{$0.1$} & \multicolumn{1}{|c|}{$0.918$} &
\multicolumn{1}{|c|}{$0.008$} & \multicolumn{1}{|c|}{$0.008$} &
\multicolumn{1}{|c|}{$0.458$} & \multicolumn{1}{|c|}{$0.073$}\\
\multicolumn{1}{|c|}{$0.3$} & $0.368$ & $0.183$ & $0.213$ & $0.733$ &
$0.043$\\
\multicolumn{1}{|c|}{$0.5$} & \multicolumn{1}{|c|}{$0.368$} &
\multicolumn{1}{|c|}{$0.183$} & \multicolumn{1}{|c|}{$0.213$} &
\multicolumn{1}{|c|}{$0.733$} & \multicolumn{1}{|c|}{$0.038$}\\
\multicolumn{1}{|c|}{$0.7$} & \multicolumn{1}{|c|}{$0.158$} &
\multicolumn{1}{|c|}{$0.358$} & \multicolumn{1}{|c|}{$0.363$} &
\multicolumn{1}{|c|}{$0.823$} & \multicolumn{1}{|c|}{$0.018$}\\
\multicolumn{1}{|c|}{$0.9$} & \multicolumn{1}{|c|}{$0.003$} &
\multicolumn{1}{|c|}{$0.753$} & \multicolumn{1}{|c|}{$0.688$} &
\multicolumn{1}{|c|}{$0.893$} & \multicolumn{1}{|c|}{$0.003$}\\\hline
\end{tabular}
\caption{Error characteristics for Males  in Example 6 at various weights.}\label{maletab2}%
\end{table}%

\section{Conclusions}

Inferences for an ROC analysis have been implemented using a characterization
of statistical evidence based on how data changes beliefs. Several contexts
have been considered, namely, a diagnostic variable taking finitely many values
with no restrictions on the distributions, a continuous diagnostic with both
distributions normal and a continuous diagnostic with no restrictions on the
distributions. A central theme is that it is not enough to simply quote the
AUC as a high value does not imply a good diagnostic. An analysis of a
diagnostic should also involve the relevant prevalence of the condition in
question as this affects the error characteristics at a specific cutoff. While
sometimes a usable optimal cutoff can be determined that takes into account
the relevant prevalence, this is not always the case and then some other
criterion needs to be considered to determine the cutoff to be used. For the
cutoff used, the error characteristics that involve the relevant prevalence
can still be assessed.

\section{Acknowledgements}

This research was supported by a grant from the Natural Sciences and
Engineering Research Council of Canada and a University of Toronto Excellence
Award. Qiaoyu Liang thanks Zhanhua He, Justin Ko, Zeyong Jin and Jiyuan Cheng
for their help.

\section*{References}

\noindent Antoniak, C. \ E. (1974) Mixtures of Dirichlet processes with
applications to Bayesian nonparametric problems. The Annals of Statistics, 2,
6, 1152 - 1174.\smallskip

\noindent Carvalho, V. de, Jara, A., Hanson, E. and Carvalho, M. de. (2013)
Bayesian nonparametric ROC regression modeling. Bayesian Analysis , 3,
623-646.\smallskip

\noindent Charvadeh, Y. K. and Yi, G. Y. (2020) Data visualization and
descriptive analysis for understanding epidemiological characteristics of
COVID-19: a case study of a dataset from January 22, 2020 to March 29, 2020.
J. of Data Sci., 18, 3.\smallskip

\noindent Doss, H. (1994) Bayesian Nonparametric Estimation for Incomplete
Data Via Successive Substitution Sampling. The Annals of Statistics, 22, 4,
1763-1786.\smallskip

\noindent Englert, B-G., Evans, M., Jang, G-H., Ng, H-K., Nott, D. and Seah
Y-L. (2018) Checking the model and the prior for the constrained multinomial.
arXiv:1804.\newline06906 and to appear in Metrika.\smallskip

\noindent Evans, M. (2015) Measuring Statistical Evidence Using Relative
Belief. Monographs on Statistics and Applied Probability 144, CRC Press,
Taylor \& Francis.\smallskip

\noindent Evans, M., Guttman, I. and Li, P. (2017) Prior elicitation,
assessment and inference with a Dirichlet prior. Entropy 2017, 19(10),
564.\smallskip

\noindent Gu, J., Ghosal, S. and Roy, A. (2008) Bayesian bootstrap estimation
of ROC\ curve. Statistics in Medicine, 27:5407--5420,
DOI10:1002/sim.3366.\smallskip

\noindent Hand, D. (2009) Measuring classifier performance: a coherent
alternative to the area under the ROC curve. Machine Learning, 99,
103-123.\smallskip

\noindent Ishwaran, H., and Zarepour, M. (2002). Exact and approximate sum
representations for the Dirichlet process. Canadian Journal of Statistics, 30,
269-283.\smallskip

\noindent Ladouceur, M., Rahme, E., Belisle, P., Scott, A., Schwartzman, K.
and Joseph, L. (2011) Modeling continuous diagnostic test data using
approximate Dirichlet process distributions. Statistics in Medicine, 30,
2648-2662.\smallskip

\noindent L\'{o}pez-Rat\'{o}n, M., Rodr\'{\i}guez-\'{A}lvarez, M. X.,
Cadarso-Su\'{a}rez, C. and Gude-Sampedro, F. (2014) OptimalCutpoints: An R
package for selecting optimal cutpoints in diagnostic tests. J. of Statistical
Software, Articles, 61, 8, doi = 10.18637/jss.v061.i08.\smallskip

\noindent Metz, C. and Pan, X. (1999) "Proper" binormal ROC curves: theory and
maximum-likelihood estimation. J. of Mathematical Psychology, 43,
1-33.\smallskip

\noindent Obuchowski, N. and Bullen, J. (2018) Receiver operating
characteristic (ROC) curves: review of methods with applications in diagnostic
medicine. Physics in Medicine \& Biology, 63, 7, 1-28.\smallskip

\noindent Swartz, T. (1993) Subjective priors for the Dirichlet process.
Communications in Statistics-Theory Methods, 28(12), 2821-2841.\smallskip

\noindent Swartz, T. (1999) Nonparametric goodness-of-fit. Communications in
Statistics-Theory Methods, 22(11), 2999-3011.\smallskip

\noindent Unal, I. (2017) Defining an optimal cut-point value in ROC
analysis:an alternative approach. Computational and Mathematical Models in
Medicine. https://doi.org/10.1155/2017/3762651.\smallskip

\noindent Verbakel, J.Y., Steyerberg, E.W., Uno, H., De Cock, B., Wynants, L.,
Collins, G.S. and Van Calster, B. (2020) ROC plots showed no added value above
the AUC when evaluating the performance of clinical prediction models. In
press, J. of Clinical Epidemiology.\smallskip

\noindent Zhou, X., Obuchowski, N. and McClish, D. (2011) Statistical Methods
in Diagnostic Medicine, 2nd Edition, Wiley.

\section*{Appendix}

\paragraph{Proof of Lemma 2}

Consider $\int_{-\infty}^{\infty}\Phi\left(  a+bz\right)  \varphi(z)\,dz$ as a
function of $b,$ so $\frac{d}{db}\int_{-\infty}^{\infty}\Phi\left(
a+bz\right)  \varphi(z)\,dz=\int_{-\infty}^{\infty}z\varphi\left(
a+bz\right)  \varphi(z)\,dz=\frac{1}{\sqrt{2\pi}\sqrt{1+b^{2}}}\exp\left(
-\frac{a^{2}}{2(1+b^{2})}\right)  $\newline$\times\int_{-\infty}^{\infty
}z\sqrt{1+b^{2}}\varphi(\sqrt{1+b^{2}}(z-(1+b^{2})^{-1}ab))\,dz=\frac{1}%
{\sqrt{2\pi}\sqrt{1+b^{2}}}\exp\left(  -\frac{a^{2}}{2(1+b^{2})}\right)
\frac{ab}{1+b^{2}}.$ When $a>0,$ then $\int_{-\infty}^{\infty}\Phi\left(
a+bz\right)  \varphi(z)\,dz$ is increasing in $b$ for $b>0$, decreasing in $b$
for $b<0,$ equals 0 when $b=0$ and when $a<0$ it is decreasing in $b$ for
$b>0$, increasing in $b$ for $b<0.$ Therefore, when $a>0,b>0,$ then
$\int_{-\infty}^{\infty}\Phi\left(  a+bz\right)  \varphi(z)\,dz\geq$
$\Phi\left(  a\right)  >1/2$ and when $a\leq0,b>0$ then $\int_{-\infty
}^{\infty}\Phi\left(  a+bz\right)  \varphi(z)\,dz\leq$ $\Phi\left(  a\right)
\leq1/2.$ $\blacksquare$

\paragraph{Proof of Lemma 3}

Note that $c_{opt}$ will satisfy $\frac{d}{dc}$Error$(c)=\frac{w}{\sigma_{D}%
}\varphi\left(  \frac{c-\mu_{D}}{\sigma_{D}}\right)  -\frac{1-w}{\sigma_{ND}%
}\varphi\left(  \frac{c-\mu_{ND}}{\sigma_{ND}}\right)  =0$ which implies
\begin{equation}
\varphi\left(  \frac{c-\mu_{D}}{\sigma_{D}}\right)  /\varphi\left(
\frac{c-\mu_{ND}}{\sigma_{ND}}\right)  =\frac{1-w}{w}\frac{\sigma_{D}}%
{\sigma_{ND}} \label{bieq4}%
\end{equation}
So $c_{opt}$ is a root of the quadratic $\left(  1/\sigma_{D}^{2}%
-1/\sigma_{ND}^{2}\right)  c^{2}-2(\mu_{D}/\sigma_{D}^{2}-\mu_{ND}/\sigma
_{ND}^{2})c+(\mu_{D}^{2}/\sigma_{D}^{2}-\mu_{ND}^{2}/\sigma_{ND}^{2}%
+2\log((1-w)\sigma_{D}/w\sigma_{ND}))$. A single real root exists when
$\sigma_{D}^{2}=\sigma_{ND}^{2}=\sigma^{2}$ and is given by (\ref{bieq1}).
When $\sigma_{D}^{2}\neq\sigma_{ND}^{2}$ there are two real roots the
discriminant $4(\mu_{D}/\sigma_{D}^{2}-\mu_{ND}/\sigma_{ND}^{2})^{2}-4\left(
1/\sigma_{D}^{2}-1/\sigma_{ND}^{2}\right)  (\mu_{D}^{2}/\sigma_{D}^{2}%
-\mu_{ND}^{2}/\sigma_{ND}^{2}+2\log((1-w)\sigma_{D}/w\sigma_{ND}))\geq0$
establishing (\ref{bieq2}). To be a minimum the root $c$ has to satisfy
$0<\frac{d^{2}\text{Error}_{w}(c)}{dc^{2}}=-\frac{w}{\sigma_{D}^{2}}\left(
\frac{c-\mu_{D}}{\sigma_{D}}\right)  \varphi\left(  \frac{c-\mu_{D}}%
{\sigma_{D}}\right)  +\frac{1-w}{\sigma_{ND}^{2}}\left(  \frac{c-\mu_{ND}%
}{\sigma_{ND}}\right)  \varphi\left(  \frac{c-\mu_{ND}}{\sigma_{ND}}\right)  $
and by (\ref{bieq4}), this holds iff $0<-\frac{w}{\sigma_{D}^{2}}\left(
\frac{c-\mu_{D}}{\sigma_{D}}\right)  \frac{1-w}{w}\frac{\sigma_{D}}%
{\sigma_{ND}}+\frac{1-w}{\sigma_{ND}^{2}}\left(  \frac{c-\mu_{ND}}{\sigma
_{ND}}\right)  =\frac{1-w}{\sigma_{ND}}\left\{  \frac{c-\mu_{ND}}{\sigma
_{ND}^{2}}-\frac{c-\mu_{D}}{\sigma_{D}^{2}}\right\}  $ which is true iff
$(1/\sigma_{D}^{2}-1/\sigma_{ND}^{2})c<\mu_{D}/\sigma_{D}^{2}-\mu_{ND}%
/\sigma_{ND}^{2}.$ When\ $\sigma_{D}^{2}=\sigma_{ND}^{2}$ this is true iff
$\mu_{D}>\mu_{ND}$ which completes the proof of (i). When $\sigma_{D}^{2}%
\neq\sigma_{ND}^{2}$ this, together with the formula for the roots of a
quadratic establishes (\ref{bieq3}). $\blacksquare$

\paragraph{Proof of Corollary 4}

Suppose $\mu_{D}\geq\mu_{ND}$ and (\ref{bieq2}) hold. Then putting $a=2\left(
\sigma_{D}^{2}-\sigma_{ND}^{2}\right)  \log((1-w)w^{-1}\sigma_{D}\sigma
_{ND}^{-1})$ we have that, for fixed $\mu_{D},\sigma_{D}^{2},\sigma_{ND}^{2}$
and $w,$ then $(\mu_{D}-\mu_{ND})^{2}+a$ is a quadratic in $\mu_{ND}.$ This
quadratic has discriminant $-4a$ and so has no real roots whenever $a>0$ and,
noting $a$ does not depend on $\mu_{D},$ the only restriction on $\mu_{ND}$ is
$\mu_{ND}\leq\mu_{D}.$ When $a\leq0$ the roots of the quadratic are given by
$\mu_{D}\pm\sqrt{-a}$ and so, since the quadratic is negative between the
roots and $\mu_{D}-\sqrt{-a}\leq\mu_{D}\leq\mu_{D}+\sqrt{-a}$ the two
restrictions imply $\mu_{ND}\leq\mu_{D}-\sqrt{-a}.$ Combining the two cases
gives (\ref{bieq3a}).

Now suppose (\ref{bieq3a}) holds. Then $\mu_{ND}\leq\mu_{D}-\{\max
(0,-a)\}^{1/2}\leq\mu_{D}\ $which gives the first restriction and also
$\mu_{ND}-\mu_{D}\leq-\{\max(0,-a)\}^{1/2}\leq0$ which implies $\left(
\mu_{ND}-\mu_{D}\right)  ^{2}\geq\max(0,-a)$ and so $\left(  \mu_{ND}-\mu
_{D}\right)  ^{2}+a\geq\max(0,-a)+a$ and by examining the cases $a\leq0$ and
$a>0\,$we conclude that (\ref{bieq2}) holds. $\blacksquare$

\end{document}